\documentstyle [psfig] {article} 

\def \sm {Standard Model }
\def \susy {supersymmetry }
\def \susyq {supersymmetric }

\def\L {\Lambda }
\def\l {\lambda }

\def \ud {{1 \over 2} }

\def \bea {\begin{equation} }
\def \eea {\end{equation} }
\def \cc {coupling constant }
\def \ccs {coupling constants }

\def \Eslash {E \kern-.5em\slash }
\def \pslash {p \kern-.5em\slash }
\def \kslash {k \kern-.5em\slash }
\newcommand{\rpv}{\mbox{$\not \hspace{-0.15cm} R_p$ }}

\title{Single chargino production at linear colliders}
\author{{G. Moreau} \\ 
{\em  Service de Physique Th\'eorique}\\
{ \em  CE-Saclay F-91191 Gif-sur-Yvette, Cedex France}}

\if@twoside\oddsidemargin25mm
\else\oddsidemargin25mm\evensidemargin25mm\marginparwidth25mm\fi
\begin{document}
\maketitle

\begin{center}  {\bf LC-TH-2000-040 }  \end{center}
\vspace{0.5cm}

\begin{abstract}
We study the single chargino production 
$e^+ e^- \to \tilde \chi^{\pm} \mu^{\mp}$  
at linear colliders which occurs through the
$\l_{121}$ R-parity violating coupling constant. 
We focus on the final state containing 4 leptons and some missing energy. 
The largest background is \susyq and 
can be reduced using the initial beam polarization and
some cuts based on the specific kinematics of the single chargino production.
Assuming the highest allowed supersymmetric background,
a center of mass energy of $\sqrt s=500GeV$ 
and a luminosity of ${\cal L}=500fb^{-1}$, the sensitivities on the
$\l_{121}$ coupling constant obtained from the single chargino production study
improve the low-energy experimental limit over a range of $\Delta m_{\tilde \nu}
\approx 500GeV$ around the sneutrino resonance, and reach values of
$\sim 10^{-4}$ at the $\tilde \nu$ pole.
The single chargino production also allows 
to reconstruct the $\tilde \chi_1^{\pm}$,
$\tilde \chi_2^{\pm}$ and $\tilde \nu$ masses.
The initial state radiation plays a fundamental role in this study.
\end{abstract}

\section{Introduction}
\label{intro}

In \susyq theories, there is no clear theoretical argument 
in favor of the conservation of the so-called 
R-parity symmetry, either from the point of view of 
grand unified models, string theories
or scenarios with discrete gauge symmetries \cite{Drein}.
The phenomenology of \susy (SUSY) at futur colliders would change
fundamentally if the R-parity symmetry were violated. 
Indeed, in such a scenario the
typical missing energy signature caused by the stable nature of the Lightest
Supersymmetric Particle (LSP) would be replaced by multijet or
multileptonic signals, depending on what are the dominant R-parity violating
(\rpv) couplings. The reason is that
the \rpv terms of the superpotential trilinear in the
quarks and leptons superfields (see Eq.(\ref{super})) allow the LSP,
whatever it is, to decay.
\begin{eqnarray}
W_{\rpv}=\sum_{i,j,k} \bigg (\ud \l _{ijk} L_iL_j E^c_k+
\l ' _{ijk} L_i Q_j D^c_k+ \ud \l '' _{ijk} U_i^cD_j^cD_k^c   \bigg ).
\label{super}
\end{eqnarray}
The effects of the \rpv decays of the LSP on the study of 
SUSY particles pair production have been considered in
the context of linear colliders \cite{Godbole}
and futur hadronic colliders, namely the Tevatron (Run II) 
\cite{Barg,tat1,tat2,runII} and the LHC \cite{Atlas}.

The measure of the \rpv coupling constants of Eq.(\ref{super}) could be 
performed via the detection of the displaced vertex associated to the decay
of the LSP. 
The sensitivities on the \rpv couplings obtained through this method
depend on the detector geometry and performances.
Let us estimate the largest values of 
the \rpv coupling constants that can be measured 
via the displaced vertex analysis. We suppose that the LSP
is the lightest neutralino ($\tilde \chi^0_1$). 
The flight length of the LSP in the laboratory frame
is then given in meters by \cite{Dreinoss}, 
\begin{eqnarray}
 c \gamma \tau \sim 3 \gamma \ 10^{-3} m ({\tilde m \over 100GeV })^4
       ({1GeV \over m_{LSP}})^5 ({1 \over \L})^2,
\label{Fleng}
\end{eqnarray}
where $\L=\l$, $\l'$ or $\l''$, 
$c$ is the light speed, $\gamma$ the Lorentz boost factor,
$\tau$ the LSP life time, $m_{LSP}$ the LSP mass and  $\tilde m$ the mass 
of the supersymmetric scalar particle 
involved in the three-body decay of the LSP.
Since the displaced vertex analysis is an experimental challenge at
hadronic colliders, we consider here the linear colliders.
Assuming that the minimum distance between two vertex necessary to distinguish 
them experimentally is of order $2 \ 10^{-5}m$ at linear colliders, we see from
Eq.(\ref{Fleng}) that the \rpv couplings could be measured up to the values,   
\begin{eqnarray}
 \L < 1.2 \ 10^{-4} \gamma^{1/2} ({\tilde m \over 100GeV })^2
       ({100GeV \over m_{LSP}})^{5/2}.
\label{LSP}
\end{eqnarray}
There is a gap between these values and the low-energy 
experimental constraints on the \rpv
couplings which range typically in the interval $\L <10^{-1}-10^{-2}$ 
for superpartners masses of $100GeV$ \cite{Drein,Bhatt,GDR}.
However, the domain lying between these low-energy bounds and the
values of Eq.(\ref{LSP}) can be tested through another way: The study of the
production of either \sm or SUSY particles involving \rpv couplings.
Indeed, the cross sections of such productions are directly proportional
to a power of the relevant \rpv coupling constant(s), which allows
to determine the values of the \rpv couplings.
Therefore, there exists a complementarity between the displaced vertex analysis
and the study of reactions involving \rpv couplings, since these
two methods allow to
investigate different ranges of values of the \rpv coupling constants.

The studies of the \rpv contributions to \sm particle productions
have been performed both at leptonic \cite{Dim1}-\cite{Deba} and hadronic
\cite{Dim2}-\cite{Rizz} colliders.
Those contributions generally involve two \rpv vertex and have thus some
rates proportional to $\Lambda^4$. The processes involving 
only one \rpv vertex are less suppressed since 
the \rpv couplings incur stringent experimental limits \cite{Drein,Bhatt,GDR}.
Those reactions correspond to the single production of \susyq particle.
Another interest of the single superpartner production 
is the possibility to produce
SUSY particles at lower energies than through the superpartner pair production, 
which is the favored reaction in R-parity conserved models.

At hadronic colliders, the single superpartner 
production involves either $\l'$ or $\l''$
\ccs \cite{Dreinoss,Dim2,Rich,Berg,greg,Rich2,Gia1,Gia2}. 
The test of the $\l$ couplings via the single superpartner 
production can only be performed  at leptonic colliders.
At leptonic colliders, either gaugino (not including gluino) 
\cite{Han,Workshop,lola,chemsin} or slepton (charged or
neutral) \cite{chemsin} can be singly produced in the simple $e^+ e^- \to
2-body$ reactions.
The single production of slepton has a reduced phase space, since the 
slepton is
produced together with a $Z$ or $W$ gauge boson. In contrast,
the $\tilde \chi^0_1$ is
produced together with a neutrino.
Nevertheless, if the $\tilde \chi^0_1$ is the LSP, as in
many \susyq scenarios, it undergoes an \rpv decay which reads as
$\tilde \chi^0_1 \to l \bar l \nu$ if one assumes a dominant $\l$ Yukawa
coupling. Therefore, the single $\tilde \chi^0_1$ production 
leads typically to the $2l + \Eslash$ final state which has a large
\sm background. In this paper, we focus on the single
chargino production
$e^+ e^- \to \tilde \chi^{\pm} l^{\mp}_m$, which occurs through
the \rpv \ccs $\l_{1m1}$ ($m=2$ or $3$).
There are several motivations. First, the single chargino production has an
higher
cross section than the single neutralino production \cite{chemsin}. Moreover, it
can lead to
particularly interesting multileptonic signatures due to the cascade decay
initiated by the chargino.

The single gaugino productions at $e^+ e^-$ colliders
have $t$ and $u$ channels
and can also receive a contribution from the resonant sneutrino production.
At $\sqrt s=200 GeV$, the off-resonance rates of the
single chargino and neutralino productions are typically of order $100 fb$
and $10 fb$, respectively \cite{chemsin},  for a
value of the relevant \rpv \cc equal to its low-enery bound for
$m_{\tilde e_R}=100GeV$: $\l_{1m1}=0.05$ \cite{Bhatt}. 
The off-pole effects of the
single gaugino production are thus at the limit of observability at LEP II
assuming an
integrated luminosity of ${\cal L } \approx 200 pb^{-1}$. At the sneutrino
resonance, the single gaugino production has higher cross section values. 
For instance with $\l_{1m1}=0.01$, the chargino production
rate can reach $2 \ 10^{-1} pb$ at the resonance \cite{Workshop}. This
is the reason why the experimental analysis of the single gaugino 
production at the LEP collider \cite{ALEPHa,DELPHIa,Fab1,Fab2,DELPHIb} allows
to test \rpv couplings values smaller than the low-energy bounds only at the
sneutrino resonance $\sqrt s = m_{\tilde \nu}$ and, due to the
Initial State Radiation (ISR) effect, in a range of typically $\sim 50GeV$
around the $\tilde \nu$ pole. The sensitivities on the $\l_{1m1}$
couplings obtained at LEP reach values of order $10^{-3}$ at the
sneutrino resonance.

The experimental analysis of the single gaugino production at futur
linear colliders should be interesting due to
the high luminosities and energies expected at these futur colliders
\cite{Tesla,NLC}. However, the single gaugino production might suffer
a large SUSY background at linear colliders. Indeed, due to the
high energies reached at these colliders, the pair productions 
of SUSY particles may have important cross sections.

In this article, we study the single chargino production via the 
$\l_{121}$ coupling at linear colliders and we consider the final 
state containing 4 leptons plus some missing energy ($\Eslash$). 
We show that the SUSY background can be
greatly reduced with respect to the signal.
This discrimination is based on the two following points:
First, the SUSY background can be suppressed by making use of the beam
polarization capability of the linear colliders.
Secondly, the specific kinematics of the single
chargino production reaction allows to put some efficient
cuts on the transverse momentum of the lepton produced together with the
chargino. We find that, by consequence of this background reduction,
the sensitivity on the $\l_{121}$ coupling obtained at the $\tilde \nu$ resonance
at linear colliders 
for $\sqrt s=500GeV$ and ${\cal L}=500 fb^{-1}$
\cite{Tesla} would be of order $10^{-4}$, namely one order of 
magnitude better than the results of the LEP analysis 
\cite{ALEPHa,DELPHIa,Fab1,Fab2,DELPHIb},
assuming the largest supersymmetric background allowed 
by the experimental limits on the SUSY masses.

Besides, in the scenario of a single dominant \rpv coupling of type $\l$
with $\tilde \chi^0_1$ as the LSP,
the experimental superpartner mass reconstruction
from the SUSY particle pair production
suffers an high combinatorial background
both at linear colliders \cite{Godbole} and LHC \cite{Atlas}.
The reason is that all the masses reconstructions are based on
the $\tilde \chi^0_1$ reconstruction which is degraded by the
imperfect identification of the charged leptons generated in the decays of the 2
neutralinos as $\tilde \chi^0_1 \to l \bar l \nu$, and the
presence of missing energy in the final state.
In particular, the chargino mass reconstruction in the leptonic channel
is difficult since the presence of an additional neutrino, coming from
the decay $\tilde \chi^{\pm} \to \tilde \chi^0 l \nu$,
renders the control on the missing energy more challenging.
In this paper, we show that
through the study of the $4l+\Eslash$ final state, the specific
kinematics of the single chargino production reaction
at linear colliders allows to determine
the $\tilde \chi^{\pm}_{1,2}$ and $\tilde \nu$ masses.

In Section \ref{frame} we define the theoretical framework.
In Sections \ref{signal} and \ref{background} we describe the signal
and the several backgrounds.
In Section \ref{An}, we present the sensitivity on the SUSY parameters
that can be obtained in an $e^+ e^-$ machine
allowing an initial beam polarization.
In this last section, we also show how some information on the SUSY mass spectrum
can be derived from the study of the single chargino production,
and we comment on another kind of signature: the $3l+2jets+\Eslash$ final state.

\section{Theoretical framework}
\label{frame}

We work within the
Minimal Supersymmetric Standard Model (MSSM) which has the
minimal particle content and the
minimal gauge group $SU(3)_C \times SU(2)_L \times U(1)_Y$,
namely the \sm gauge symmetry.
The \susyq parameters defined at the electroweak scale are
the Higgsino mixing parameter $\mu$,
the ratio of the vacuum expectation values of the two Higgs doublet
fields $\tan \beta=<~H_u~>~/~<~H_d~>$ and the soft SUSY breaking parameters,
namely the bino ($\tilde B$) mass $M_1$,
the wino ($\tilde W$) mass $M_2$, the gluino ($\tilde g$) mass $M_3$,
the sfermion masses $m_{\tilde f}$ and the trilinear Yukawa couplings $A$.
The remaining three
parameters $m_{H_u}^2$, $m_{H_d}^2$ and the
soft SUSY breaking bilinear coupling $B$ are
determined through the electroweak symmetry breaking conditions,
which are two necessary minimization conditions of the Higgs potential.

We assume that all phases in the soft SUSY breaking potential are equal to
zero in order to eliminate all new sources of CP violation which are
constrained by extremely tight experimental limits on the electron and
neutron electric moments. Furthermore, to avoid any problem of
Flavor Changing Neutral Currents, we take the matrices in flavor space
of the sfermion masses and $A$ couplings
close to the unit matrix. In particular, for simplification reason
we consider vanishing $A$ couplings.
This last assumption concerns the splitting
between the Left and Right sfermions masses
and does not affect our analysis which depends mainly on the relative
values of the sleptons, squarks and gauginos masses as we will discuss
in next sections.

Besides, we suppose the R-parity symmetry to be violated so that
the interactions written in the superpotential of Eq.(\ref{super})
are present. 
The existence of a hierarchy among the values of the \rpv couplings
is suggested by the analogy of those couplings
with the Yukawa couplings.
We thus assume a single dominant \rpv coupling of the type $\l_{1m1}$
which allows the single chargino production at $e^+e^-$ colliders.

Finally, we choose the LSP to be the $\tilde\chi^0_1$ neutralino
since it is often real in many models, such as the supergravity
inspired models.

\section{Signal}
\label{signal}

\begin{figure}[t]
\begin{center}
\leavevmode
\centerline{\psfig{figure=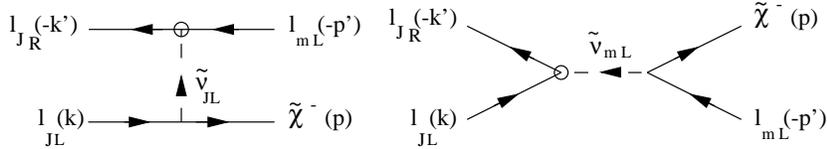}}
\end{center}
\caption{\footnotesize  \it
Feynman diagrams of the single chargino production process at
leptonic colliders.
The \rpv coupling constant $\l_{JmJ}$
($J=1$, $m=2 \ or \ 3$ for $e^+ e^-$ colliders and
$J=2$, $m=1 \ or \ 3$ for $\mu^+ \mu^-$ colliders)
is symbolised by a small circle,
and the arrows denote the flow of momentum of the corresponding particles.
The Left and Right helicities of the particles are denoted by L and R,
respectively, and the momentum of the particles by k,k',p and p'.
The higgsino contribution as well as the charge conjugated process
are not represented.
\rm \normalsize }
\label{graphe}
\end{figure}

The single chargino production $e^+ e^- \to \tilde \chi^{\pm} l^{\mp}_m$
occurs via the $\l_{1m1}$ coupling constant 
either through the exchange of a $\tilde \nu_{mL}$ sneutrino in the $s$
channel or a $\tilde \nu_{eL}$ sneutrino in the $t$ channel (see
Fig.\ref{graphe}).
Due to the antisymmetry of the $\l_{ijk}$ Yukawa couplings in the $i$ and
$j$ indices,
only a muon or a tau-lepton can be produced together with the chargino,
namely $m=2$ or $3$.
The produced chargino can decay into the LSP as
$\tilde \chi^{\pm} \to \tilde \chi^0_1  l \nu$
or as $\tilde \chi^{\pm} \to \tilde \chi^0_1 u d$.
Then the LSP decays as $ \tilde \chi_1^{0} \to  \bar e e \nu_m$,
$\bar e e \bar \nu_m$, $l_m \bar e \bar \nu_e$ 
or $ \bar l_m e \nu_e $ through $\l_{1m1}$. 
Each of these 4 $ \tilde \chi_1^{0}$ decay channels   
has a branching ratio exactly equal to $25\%$.
In this paper,
we study the $4l + \Eslash$ final state generated by the decay
$\tilde \chi^{\pm} \to \tilde \chi^0_1  l \nu$ of the produced chargino.

The single chargino production has a specific feature which
is particularly attractive: Because of 
the simple kinematics of $2 \to 2-body$ type reaction,
the energy of the lepton produced with the chargino, which we write $E(l_m)$,
is completely fixed by the center of mass energy $\sqrt s$,
the lepton mass $m_{l_m}$ and the chargino
mass $m_{\tilde \chi^{\pm}}$:
\begin{eqnarray}
E(l_m)= {s + m^2_{l_m} - m^2_{\tilde \chi^{\pm}} \over 2 \sqrt s}.
\label{enl}
\end{eqnarray}
The lepton momentum $P(l_m)$, which is related to the lepton energy by 
$P(l_m)=(E(l_m)^2-m^2_{l_m}c^4)^{1/2}/c$, is thus also fixed. 
Therefore, the momentum distribution 
of the produced lepton is peaked and offers the opportunity to
put some cuts allowing an effective discrimination between the
signal and the several backgrounds. Besides, the momentum value of the
produced lepton gives the $\tilde \chi^{\pm}$ mass through Eq.(\ref{enl}).
Nevertheless, for a significant ISR effect the radiated photon 
carries a non negligible energy and the single chargino production 
must be treated as a three-body reaction of the type 
$e^+ e^- \to \tilde \chi^{\pm} l^{\mp}_m \gamma$. Therefore,
the energy of the produced lepton is not fixed anymore by the SUSY parameters.
As we will show in Section \ref{kin}, in this situation the transverse
momentum of the produced lepton is more appropriate to apply some cuts
and reconstruct the $\tilde \chi^{\pm}$ mass.
\\ In the case where the produced lepton is a tau, the momentum of the lepton 
($e^{\pm}$ or $\mu^{\pm}$) coming from the $\tau$-decay is of
course not peaked at a given value. Moreover, 
in contrast with the muon momentum, the precise determination
of the tau-lepton momentum is difficult experimentally  
due to the unstable nature of the $\tau$.
From this point of view, the case of a single dominant $\l_{121}$
coupling constant is a more promising scenario
than the situation in which $\l_{131}$ is the dominant coupling.
Hence, from now on we focus on the single dominant \rpv coupling
$\l_{121}$ and consider the single chargino production 
$e^+ e^- \to \tilde \chi^{\pm} \mu^{\mp}$.

At this stage, an important remark can be made concerning the
initial state.
As can be seen from Fig.\ref{graphe}, in the single $\tilde \chi^-$
production
the initial electron and positron have the same helicity, namely, they are
both Left-handed.
Similarly in the $\tilde \chi^+$ production, the electron and the positron
are
both Right-handed. The incoming lepton and anti-lepton 
have thus in any case the same helicity.
This is due to the particular structure of the trilinear \rpv couplings
which flips the chirality of the fermion. 
The incoming electron and positron could also have identical helicities.
However, this contribution involves the higgsino component of the chargino
and is thus suppressed by the coupling of the higgsino to leptons which
is proportional to $m_l / (m_W \ \cos \beta )$.
This same helicity property, which is characteristic
of all type of single superpartner
production at leptonic colliders \cite{chemsin},
allows to increase the single chargino production rate by selecting
same helicities initial leptons.

\section{Background}
\label{background}

\subsection{Non-physic background}
\label{NPback}

\begin{figure}[t]
\begin{center}
\leavevmode
\centerline{\psfig{figure=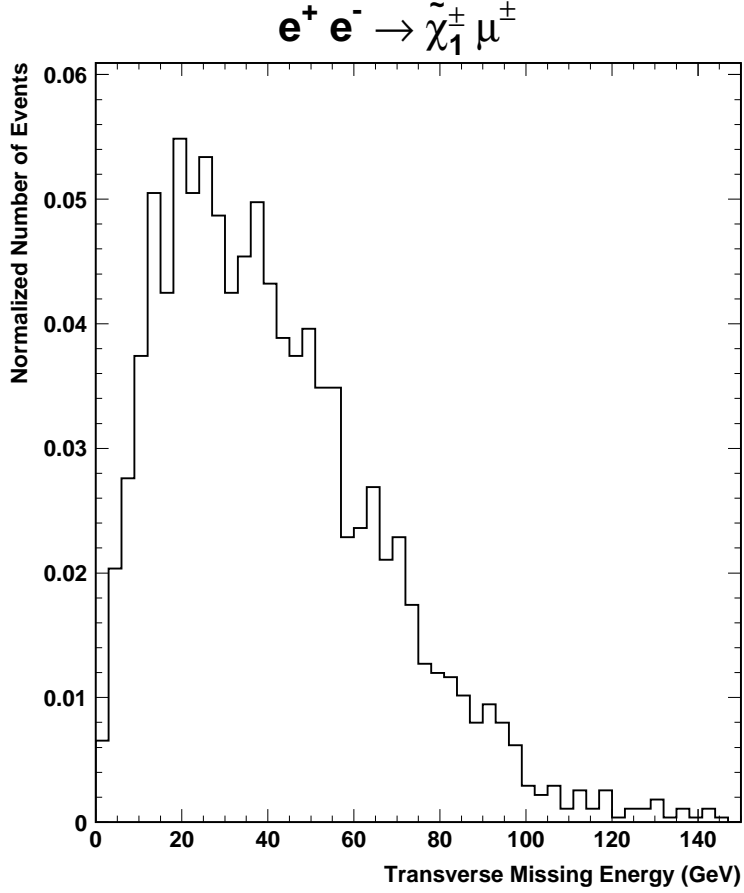,height=5.in}}
\end{center}
\caption{\footnotesize  \it
Distribution of the transverse missing energy (in $GeV$)
in the $4l+\Eslash$ events generated
by the single $\tilde \chi^{\pm}_1$ production 
at a center of mass energy of $500GeV$ and
for the point A of the SUSY parameter
space with $m_{\tilde \nu}=240GeV$.
The number of events has been normalized to unity.
\rm \normalsize }
\label{distrib}
\end{figure}

First, one has to consider the non-physic background for the $4l +
\Eslash$ signature.
The main source of such a background is the $Z$-boson pair production with
ISR.
Indeed, the ISR photons have an high probability
to be colinear
to the beam axis. They will thus often be missed experimentally becoming
then
a source of missing energy. The four leptons can come from the leptonic
decays of
the 2 $Z$-bosons.
This background can be greatly reduced by excluding the same flavor
opposite-sign dileptons
which have an invariant mass in the range, $10GeV < \vert M_{inv}(l_p \bar
l_p)- M_Z \vert$, $p=1,2,3$ being the generation indice.
Furthermore, the missing energy coming from the ISR is mainly
present at small angles
with respect to the beam axis. This point can be exploited to perform a
better selection of
the signal. As a matter of fact, the missing energy
coming from the signal has a significant transverse component since it
is caused by the presence of neutrinos in the final state.
In Fig.\ref{distrib}, we present the distribution of the transverse
missing energy
in the $4l +\Eslash$ events generated by the single $\tilde \chi^{\pm}_1$
production for a sneutrino mass of $m_{\tilde \nu} = 240 GeV$ and for the 
point A of the SUSY parameter space defined as:
$M_1=200GeV$, $M_2=250GeV$, $\mu=150GeV$, $\tan \beta=3$,
$m_{\tilde l^{\pm}}=300GeV$ and $m_{\tilde q}=600GeV$
($m_{\tilde \chi^{\pm}_1}=115.7GeV$, $m_{\tilde
\chi^0_1}=101.9GeV$, $m_{\tilde \chi^0_2}=154.5GeV$).
The cut on the lepton momentum mentioned in Section \ref{signal} can also
enhance the signal-to-background ratio. Finally, the beam
polarization may be useful in reducing this source of background. 
Indeed, the initial leptons in the
$ZZ$ production process have opposite helicities. The reason is that the
$Z-f-\bar f$
vertex conserves the fermion chirality. Furthermore, recall that in our signal, the
initial electron and positron have similar helicities. Thus if the beam
polarization at
linear collider were chosen in order to favor opposite helicities initial states,
the signal would be enhanced while 
the $ZZ$ background would be suppressed. Since the polarization
expected at linear colliders is of order
$85 \%$ for the electron and $60 \%$ for the positron \cite{Tesla}, the
$ZZ$ background would not be entirely eliminated by the beam
polarization effect.

\subsection{\sm background}
\label{SMback}

The \sm backgrounds for the $4l + \Eslash$ signal come from $2 \to 3-body$
processes.
For example, the $e^+ e^- \to Z^0 Z^0 Z^0$ reaction can give rise to the
four leptons plus
missing energy signature. Another source of \sm background is the $e^+ e^-
\to Z^0 W^+ W^-$ reaction.
The rates for those backgrounds have been calculated in \cite{Godbole} after
the cuts on the charged lepton rapidity, $\vert \eta (l)\vert<3$,
the charged lepton transverse momentum, $P_t(l)>10GeV$, and
the missing transverse energy, $\Eslash_t>20GeV$, have been applied.
The results are the followings.
The $Z Z Z$ production cross section does not
exceed $1fb$. The $Z W W$ production has a cross section
of $0.4fb$ ($0.1fb$) at $\sqrt s=500GeV$ ($350GeV$) after convoluting
with the branching fractions for the leptonic decays of the gauge bosons.
This rate value is obtained without the ISR effect which should reduce the
cross section of the $Z W W$ production dominated
by the $t$ channel exchange contribution \cite{Godbole}.
As before, the $Z W W$ production can be reduced using simultaneously
the cut on the lepton momentum indicated in Section \ref{signal},
the dilepton invariant mass cut and the beam polarization effect. Indeed, 
the initial leptons have opposite helicities in both the $s$ and $t$ channels
of the process $e^+ e^- \to Z^0 W^+ W^-$, as in the $ZZ$ production process.

\subsection{Supersymmetric background}
\label{SUSYB}

The main sources of \susyq background to the $4l+
\Eslash$ signature are the neutralinos and sneutrinos pair productions.

First, the reaction,
\begin{eqnarray}
e^+ e^- \to \tilde \chi^0_1 + \tilde \chi^0_1,
\label{chi1}
\end{eqnarray}
represents a background for the $4l+\Eslash$ signature since
the $\tilde \chi^0_1$, which is the LSP in our
framework, decays as $ \tilde \chi_1^{0} \to  \bar e e \nu_{\mu}$,
$\bar e e \bar \nu_{\mu}$, $\mu \bar e \bar \nu_e$ 
or $ \bar \mu e \nu_e $ 
through the $\l_{121}$ coupling. Therefore, the
$\tilde \chi^0_i \tilde \chi^0_j$  productions
(where $i=1,...,4$ and $j=1,...,4$ are not equal to $1$ simultaneously)
leading through cascade decays
to a pair of $\tilde \chi^0_1$ accompanied by some
neutrinos give also rise to the final state
with 4 leptons plus some missing energy.
These reactions are the following,
\begin{eqnarray}
e^+ e^- \to \tilde \chi^0_i + \tilde \chi^0_1 \to \tilde \chi^0_1 \nu_p \bar
\nu_p + \tilde \chi^0_1,
\label{chi2}
\end{eqnarray}
\begin{eqnarray}
e^+ e^- \to \tilde \chi^0_i + \tilde \chi^0_j \to \tilde \chi^0_1 \nu_p \bar
\nu_p + \tilde \chi^0_1 \nu_{p'} \bar \nu_{p'},
\label{chi3}
\end{eqnarray}
where $i,j=2,3,4$ and $p,p'=1,2,3$.
Besides, the pair productions
of neutralinos followed
by their \rpv decays through $\l_{121}$:
\begin{eqnarray}
e^+ e^- \to \tilde \chi^0_i + \tilde \chi^0_1 \to  l  \bar l  \nu  + \tilde
\chi^0_1,
\label{chrp1}
\end{eqnarray}
\begin{eqnarray}
e^+ e^- \to \tilde \chi^0_i + \tilde \chi^0_j \to  l  \bar l  \nu +  \tilde
\chi^0_1 \nu_p \bar
\nu_p,
\label{chrp2}
\end{eqnarray}
\begin{eqnarray}
e^+ e^- \to \tilde \chi^0_i + \tilde \chi^0_j \to  l  \bar l  \nu  + l  \bar
l  \nu,
\label{chrp3}
\end{eqnarray}
where $i,j=2,3,4$, are also a source of background.

Secondly, the $4l+\Eslash$ signature can arise
in the sneutrino pair production through the reactions,
\begin{eqnarray}
e^+ e^- \to \tilde \nu_p + \tilde \nu_p^* \to \tilde \chi^0_i \nu_p + \tilde
\chi^0_j \bar \nu_p,
\label{sneu}
\end{eqnarray}
where $i,j=1,...,4$, if the produced neutralinos decay as above,
namely either as 
$\tilde \chi^0_i \to \tilde \chi^0_1 \nu_p \bar \nu_p$ (if $i \neq 1$) or as 
$\tilde \chi^0_i \to l  \bar l  \nu$.

The neutralinos and sneutrinos pair productions also lead to the
$4l+\Eslash$ final state via more complex cascade decays such as
$\tilde \chi^0_i \to \tilde \chi^0_j \nu_p \bar \nu_p \to
\tilde \chi^0_1 \nu_{p'} \bar \nu_{p'}  \nu_p \bar \nu_p$ ($i,j=2,3,4$ and
$i>j$) or $\tilde \nu \to l^{\mp} \tilde \chi^{\pm}_1 \to l^{\mp} e^{\pm}
\nu_e \nu_{\mu}$
where the chargino decays through the $\l_{121}$ coupling.

At the high energies available at linear colliders \cite{Tesla,NLC},
the rates of the neutralinos and sneutrinos pair productions (see Section
\ref{SUSYX})
are typically quite larger than the cross sections of the \sm background
reactions (see Section \ref{SMback}).
We however note that in the regions of large $\tilde \chi^0_i$ and
$\tilde \nu_p$ masses
with respect to the center of mass energy $\sqrt s$,
the $\tilde \chi^0_i \tilde \chi^0_j$ and
$\tilde \nu_p \tilde \nu_p^*$ productions cross sections, respectively,
are
suppressed by the phase space factor.
In the kinematic domain $2m_{\tilde \chi^0_i},2m_{\tilde \nu_p}>\sqrt
s>m_{\tilde
\chi^{\pm}_1}>m_{\tilde \chi^0_1}$ ($i=1,...,4$ and $p=1,2,3$), all
the neutralinos and sneutrinos pair productions are even kinematically
closed while the
single chargino production remains possible.
In such a situation, the only background to the 
$4l+\Eslash$ final state would be the
\sm background so that the sensitivity on the $\lambda_{121}$ coupling would
be greatly improved. Nevertheless, the kinematic domain described above is
restrictive
and not particularly motivated. As a conclusion, except in some particular
kinematic domains,
the main background to the $4l+\Eslash$ final state is supersymmetric.\\
In Section \ref{An}, we will focus on the largest allowed
SUSY background.
Indeed, the various considered domains of the SUSY parameter
space will be chosen such that the $\tilde \chi^0_1$
mass is around the value
$m_{\tilde \chi^0_1} \approx 100GeV$ which is close to the experimental limit
$m_{\tilde \chi^0_1}>52GeV$ (for $\tan \beta =20$ and any $\l_{ijk}$
coupling)
\cite{DELPHIb}
in order to maximize the phase space factors of the 
$\tilde \chi^0_i \tilde
\chi^0_j$ productions and the decays $\tilde \nu_p \to 
\tilde \chi^0_i \nu_p$.

The SUSY background could be suppressed by the cut on the lepton momentum 
mentioned in Section \ref{signal}. 
The selection of initial leptons of same helicities 
would also reduce the SUSY background. Indeed,
the neutralino (sneutrino) pair production occurs through the
exchange of either a
$Z$-boson in the $s$ channel or a charged slepton (chargino) in the $t$
channel, and
the helicities of the initial electron and positron are opposite in all
those channels. The $t$ channels of the SUSY background processes 
allow however same helicity leptons in the initial state but this
contribution entails the higgsino component of the relevant gaugino
and is thus suppressed by powers of $m_l /( m_W \ \cos \beta )$.

\section{Analysis}
\label{An}

In this part, we determine the sensitivity on the $\l_{121}$
\cc expected from the study of the single chargino production
$e^+ e^- \to \tilde \chi_1^{\pm} \mu^{\mp}$ based on
the analysis of the $4l+\Eslash$ final state 
at linear colliders, assuming a center of mass energy of 
$\sqrt s=500GeV$ and a luminosity of ${\cal L}=500fb^{-1}$.
For this purpose, we study the SUSY background and show how it
can be reduced with respect to the signal. At the end of this part, 
we also discuss the determination of the 
$\tilde \chi^{\pm}_1$, $\tilde \chi^{\pm}_2$ and $\tilde \nu$ masses,
the single chargino production at different center of mass energies
and via other \rpv couplings,
the $3l+2j+\Eslash$ final state and the single neutralino production.

We have simulated the signal and the \susyq background with
the new version of the SUSYGEN event generator \cite{SUSYGEN3}
including the beam polarization effects.

\subsection{Polarization}
\label{polar}

The signal-to-noise ratio could be enhanced by 
making use of the capability of the linear colliders to polarize the
inital beams. 

First, as we have seen in Section \ref{SUSYB} the SUSY
background would be reduced if initial
leptons of similar helicities were selected.  
Hence, we consider in this study the selection 
of the $e^-_L e^+_L$ initial state, 
namely Left-handed initial electron and positron.
In order to illustrate the effect of this 
polarization on the SUSY background, let us
consider for example the $\tilde \chi^0_1 \tilde \chi^0_2$ production.
At the point A of the SUSY parameter space, the unpolarized cross section
is $\sigma(e^+ e^- \to \tilde \chi^0_1 \tilde \chi^0_2)=153.99fb$
for a center of mass energy of $\sqrt s=500GeV$.
Selecting now the $e^-_L e^+_L$ initial beams 
and assuming a value of $85 \%$  ($60 \%$)
for the electron (positron) polarization, the
production rate becomes
$\sigma_{polar}(e^+ e^- \to \tilde \chi^0_1 \tilde \chi^0_2 )=82.56fb$.

Secondly, the selection of same helicities initial
leptons would increase the signal rate as mentioned in Section \ref{signal}.
We discuss here the effect of the considered
$e^-_L e^+_L$ beam polarization on the signal rate.
As explained in Section \ref{signal}, the single chargino production
occurs through the process
$e^-_L e^+_L \to \tilde \chi^-_1 \mu^+$ (see Fig.\ref{graphe}) or through
its charge conjugated process
$e^-_R e^+_R \to \tilde \chi^+_1 \mu^-$.
These charge conjugated processes have the same cross section:
$\sigma(e^-_L e^+_L \to \tilde \chi^-_1 \mu^+)=\sigma(e^-_R e^+_R \to \tilde
\chi^+_1 \mu^-)$.
The whole cross section is thus
$\sigma=[\sigma(e^-_L e^+_L \to \tilde \chi^-_1 \mu^+)
+\sigma(e^-_R e^+_R \to \tilde \chi^+_1 \mu^-)] / 4
=\sigma(e^-_L e^+_L \to \tilde \chi^-_1 \mu^+) / 2$.
The chosen beam polarization favoring Left-handed initial leptons,
only the process $e^-_L e^+_L \to \tilde \chi^-_1 \mu^+$
would be selected if the polarization were total.
The polarized cross section would then be $\sigma_{polar}^{total}=\sigma(e^-_L e^+_L
\to \tilde \chi^-_1 \mu^+)=2 \ \sigma$. Hence,
the signal would be increased by a factor of $2$
with respect to the whole cross section.
In fact, due to the limited expected efficiency of the beam polarization,
the chargino production can also receive a small contribution from the
process
$e^-_R e^+_R \to \tilde \chi^+_1 \mu^-$.
The polarized cross section is thus replaced by
$\sigma_{polar}=\sigma_{polar}(\tilde \chi^-_1 \mu^+)+\sigma_{polar}(\tilde
\chi^+_1 \mu^-)$
where $\sigma_{polar}(\tilde \chi^-_1 \mu^+)=P(e^-_L)P(e^+_L)\sigma(e^-_L
e^+_L \to \tilde \chi^-_1 \mu^+)$
and $\sigma_{polar}(\tilde \chi^+_1 \mu^-)=P(e^-_R)P(e^+_R)\sigma(e^-_R
e^+_R \to \tilde \chi^+_1 \mu^-)$,
$P(e^+_L)$ being for example the probability to have a Left-handed initial
positron.
Assuming a polarization efficiency of $85 \%$  ($60 \%$) for the electron
(positron)
and selecting Left-handed initial leptons,
the probabilities are $P(e^-_L)=(1+0.85) / 2=0.925$, $P(e^+_L)=(1+0.60)
/ 2=0.8$,
$P(e^-_R)=(1-0.85)/ 2=0.075$ and $P(e^+_R)=(1-0.60)/ 2=0.2$. Hence,
the polarized cross section reads as
$\sigma_{polar}=[P(e^-_L)P(e^+_L)+P(e^-_R)P(e^+_R)]\sigma(e^-_L e^+_L \to
\tilde \chi^-_1 \mu^+)=
0.755 \ \sigma(e^-_L e^+_L \to \tilde \chi^-_1 \mu^+)=1.51 \ \sigma$.
The signal rate is thus enhanced by a factor of $1.51$
with respect to the whole cross section if the 
$e^-_L e^+_L$ beam polarization is applied.

\subsection{Cross sections and branching ratios}

\subsubsection{Signal}
\label{signalX}

The single $\tilde \chi^{\pm}_1$ chargino production has a cross section
typically of order $10fb$ at $\sqrt s=500GeV$ for $\l_{121}=0.05$
\cite{chemsin}. 
The single $\tilde \chi^{\pm}_1$ production rate is reduced in 
the higgsino dominated region
$\vert \mu \vert \ll M_1,M_2$ where the $\tilde \chi^{\pm}_1$ is
dominated by its higgsino component, compared to the wino
dominated domain
$\vert \mu \vert \gg M_1,M_2$ in which the $\tilde \chi^{\pm}_1$ is
mainly composed by the charged higgsino \cite{chemsin}. Besides, 
the single $\tilde \chi^{\pm}_1$ production cross section 
depends weakly on the sign of the $\mu$ parameter 
at large values of $\tan \beta$. However, as $\tan \beta$ decreases
the rate increases (decreases) for $sign(\mu)>0$ ($<0$).
This evolution of the rate 
with the $\tan \beta$ and $sign(\mu)$ parameters is 
explained by the evolution of the $\tilde \chi^{\pm}_1$ mass 
in the SUSY parameter space \cite{chemsin}.
Finally, when the sneutrino mass approaches
the resonance ($m_{\tilde \nu} =\sqrt s$) by lower values,
the single $\tilde \chi^{\pm}_1$ production
cross section is considerably increased due to the ISR effect \cite{lola}.
For instance, at the point A of the SUSY parameter space, the rate is
equal to
$\sigma (e^+ e^- \to \tilde \chi_1^{\pm} \mu^{\mp})=353.90fb$ 
for a center of mass energy of
$\sqrt s=500GeV$ and a sneutrino mass of $m_{\tilde \nu} = 240 GeV$.
At the resonance, the single chargino production rate reaches high values.
For example at $m_{\tilde \nu} =\sqrt s=500GeV$, the cross section is  
$\sigma (e^+ e^- \to \tilde \chi_1^{\pm} \mu^{\mp})=30.236pb$ 
for the same MSSM point A.

Since we consider the $4l+ \Eslash$ final state and the chargino
width is neglected,
the single chargino production rate must be multiplied by
the branching ratio of the leptonic chargino decay
$B(\tilde \chi_1^{\pm} \to \tilde \chi^0_1  l_p \nu_p)$ ($p=1,2,3$).
The leptonic decay of the chargino is typically of order $30\%$
for $m_{\tilde \nu},m_{\tilde l^{\pm}},
m_{\tilde q}>m_{\tilde \chi^{\pm}_1}$.
This leptonic decay is suppressed compared to the hadronic decay
$\tilde \chi_1^{\pm} \to \tilde \chi^0_1 d_p u_{p'}$ ($p=1,2,3;p'=1,2$)
because of the color factor.
Indeed for $m_{\tilde \nu},m_{\tilde l^{\pm}},m_{\tilde q}
>m_{\tilde \chi^{\pm}_1}$
the hadronic decay is typically
$B(\tilde \chi_1^{\pm} \to \tilde \chi^0_1 d_p u_{p'}) \approx 70\%$
($p=1,2,3;p'=1,2$).
In the case where $m_{\tilde q}>m_{\tilde \chi^{\pm}_1}>
m_{\tilde \nu},m_{\tilde l^{\pm}}$,
the decay $\tilde \chi_1^{\pm} \to \tilde \chi^0_1  l_p \nu_p$
occurs through the two-body decays
$\tilde \chi_1^{\pm} \to \tilde \nu l^{\pm}$ and
$\tilde \chi_1^{\pm} \to \tilde l^{\pm} \nu$
and is thus the dominant channel. In such a scenario,
the branching ratio of the $4l+ \Eslash$ final state 
for the single $\tilde \chi_1^{\pm}$ production is close to
$100\%$. In contrast, for
$m_{\tilde \nu},m_{\tilde l^{\pm}}>m_{\tilde \chi^{\pm}_1}>
m_{\tilde q}$, the decay
$\tilde \chi_1^{\pm} \to \tilde \chi^0_1 d_p u_{p'}$ dominates,
as it occurs through the two-body decay
$\tilde \chi_1^{\pm} \to \tilde q q'$, and the signal is
negligible. The \rpv decays of the chargino via $\l_{121}$, 
$\tilde \chi_1^{\pm} \to e^{\pm} \mu^{\pm}
e^{\mp}, \ e^{\pm} \nu_e \nu_{\mu}$,
have generally negligible branching fractions due to the
small values of the \rpv coupling constants.
Nevertheless, in the case of nearly degenerate $\tilde \chi^{\pm}_1$
and $\tilde \chi^0_1$
masses, those \rpv decays can become dominant spoiling then
the $4l+ \Eslash$ signature.

\subsubsection{SUSY background}
\label{SUSYX}

In this part, we discuss the variations and the order of magnitude of the
cross sections and branching ratios on which the whole SUSY background
depends.

$\bullet$ {\bf Neutralino pair production}
The neutralinos pair productions represent a SUSY background
for the present study of the R-parity violation.
A detailed description of the neutralinos pair productions cross sections at
linear colliders
for an energy of $\sqrt s=500GeV$ has recently been performed in
\cite{Godbole}.
In order to consider in our analysis the main variations of the neutralinos
pair productions rates,
we have generated all the $\tilde \chi^0_i \tilde \chi^0_j$ productions
($i$ and $j$ both varying between $1$ and $4$)
at some points belonging to characteristic regions of the MSSM parameter
space.
The points chosen for the analysis respect the experimental limits
derived from the LEP data on the lightest chargino and neutralino masses,
$m_{\tilde \chi^0_1}>52GeV$ ($\tan \beta=20$)
and $m_{\tilde \chi^{\pm}_1}>94GeV$  \cite{DELPHIb},
as well as the excluded regions in the $\mu - M_2$ plane \cite{DELPHIb}.
Besides, since the neutralinos pair productions 
rates depend weakly on $\tan \beta$ and $sign(\mu)$ \cite{Godbole},
we have fixed those SUSY parameters at $\tan \beta=3$ and $sign(\mu)>0$.

We present now the characteristic domains of the MSSM parameter space
considered in our analysis. For each of those
domains, we will describe the behaviour of all the
$\tilde \chi^0_i \tilde \chi^0_j$
productions cross sections except the $\tilde \chi^0_i \tilde \chi^0_4$
productions ($i=1,...,4$).
This is justified by the fact that
at a center of mass energy of $500GeV$,
the dominant neutralinos productions are the $\tilde \chi^0_1 \tilde
\chi^0_1$,
$\tilde \chi^0_1 \tilde \chi^0_2$ and $\tilde \chi^0_2 \tilde \chi^0_2$
productions
in most parts of the SUSY parameter space.
We will also discuss in each of the considered regions
the values of the branching ratios $B(\tilde
\chi^0_2 \to \tilde \chi^0_1 \nu_p \bar
\nu_p)$ and $B(\tilde \chi^0_2 \to l \bar l \nu)$ 
(for $\l_{121}=0.05$) which determine the contribution of the
$\tilde \chi^0_1 \tilde \chi^0_2$ and $\tilde \chi^0_2 \tilde \chi^0_2$ productions 
to the $4l+ \Eslash$ signature.
The restriction of the discussion to the 
$\tilde \chi^0_2$ decays is justified by the hierarchy
mentioned above between the $\tilde \chi^0_i \tilde \chi^0_j$ productions rates,
and by the fact that the $\tilde \chi^0_3$ and $\tilde \chi^0_4$ cascade decays 
have many possible combinations,
due to the large $\tilde \chi^0_3$ and $\tilde \chi^0_4$ masses
with respect to the $\tilde \chi^0_1$ and $\tilde \chi^0_2$ masses, so that
the contributions of the $\tilde \chi^0_i \tilde \chi^0_j$ productions
(with $i$ or $j$ equal to $3$ or $4$) to the $4l+\Eslash$ signature
are suppressed by small branching ratio factors.
We will also not discuss the complex cascade decays mentioned in
Section \ref{SUSYB} since the associated branching ratios are typically
small.
However, all the $\tilde \chi^0_i \tilde \chi^0_j$ productions as well as
all the cascade decays of the four neutralinos
are taken into account in the analysis.

First, we consider the higgsino dominated region characterised by
$\vert \mu \vert \ll M_1,M_2$ where the $\tilde \chi^0_1$
and $\tilde \chi^0_2$ neutralinos are predominantly composed by the higgsinos.
In this higgsino region, due to the weak couplings of the higgsinos to
charged leptons,
the $\tilde \chi^0_1 \tilde \chi^0_i$
and $\tilde \chi^0_2 \tilde \chi^0_i$ productions ($i=1,2,3$) are reduced.
However,
the $\tilde \chi^0_1 \tilde \chi^0_2$ production rate reaches its larger
values in this domain because of the $Z \tilde \chi^0_i \tilde \chi^0_j$
coupling.
At $\sqrt s=500GeV$, the
$\tilde \chi^0_3 \tilde \chi^0_3$
production is suppressed by a small phase space factor (when
kinematically allowed).
For the MSSM point A which belongs to this higgsino region,
the cross sections including the beam polarization described in Section
\ref{polar} are
$\sigma(\tilde \chi^0_1 \tilde \chi^0_1)=0.29fb$,
$\sigma(\tilde \chi^0_1 \tilde \chi^0_2)=82.56fb$,
$\sigma(\tilde \chi^0_2 \tilde \chi^0_2)=0.11fb$,
$\sigma(\tilde \chi^0_1 \tilde \chi^0_3)=6.62fb$,
$\sigma(\tilde \chi^0_2 \tilde \chi^0_3)=6.31fb$,
$\sigma(\tilde \chi^0_3 \tilde \chi^0_3)=13.57fb$,
$\sigma(\tilde \chi^0_1 \tilde \chi^0_4)=1.11fb$
and $\sigma(\tilde \chi^0_2 \tilde \chi^0_4)=9.59fb$
for the kinematically open neutralinos productions at $\sqrt s=500GeV$.\\
The $\tilde \chi^0_1 \tilde \chi^0_2$ production gives rise to the $4l+
\Eslash$ final state
if the second lightest neutralino decays as $\tilde \chi^0_2 \to \tilde
\chi^0_1 \bar \nu_p \nu_p$.
In the higgsino region, due to the various decay channels of the $\tilde
\chi^0_2$, the branching ratio
$B(\tilde \chi^0_2 \to \tilde \chi^0_1 \bar \nu_p \nu_p)$ reaches values
only about $15 \%$.
At the point A of the MSSM parameter space, this branching is
$B(\tilde \chi^0_2 \to \tilde \chi^0_1 \bar \nu_p \nu_p)=15.8\%$
for $m_{\tilde \nu}=450GeV$.
For a $\tilde \nu$ lighter than the $\tilde \chi^0_2$, the branching
$B( \tilde \chi^0_2\to \tilde \chi^0_1 \bar \nu_p \nu_p)$
is enhanced as this decay occurs via
the two-body decay $\tilde \chi^0_2 \to \tilde \nu \nu$.
At the point A with $m_{\tilde \nu}=125GeV$, 
$B( \tilde \chi^0_2\to \tilde \chi^0_1 \bar \nu_p \nu_p)=62.3\%$.

Secondly, we consider the
wino dominated region $\vert \mu \vert \gg M_1,M_2$ where $\tilde \chi^0_1$ and
$\tilde \chi^0_2$ are primarily bino and wino, respectively.
In this wino region, the $\tilde \chi^0_3 \tilde \chi^0_i$ productions
are reduced since the
$\tilde \chi^0_3$ mass strongly increases with the absolute value of the
parameter
$\vert \mu \vert $. Therefore in the wino region, the main neutralinos
productions are
the $\tilde \chi^0_1 \tilde \chi^0_1$, $\tilde \chi^0_1 \tilde \chi^0_2$ and
$\tilde \chi^0_2 \tilde \chi^0_2$
productions. In this part of the MSSM parameter space
and for $m_{\tilde l^{\pm}}=300GeV$, the $\tilde \chi^0_1
\tilde \chi^0_1$
production can reach $\sim 30fb$, after the beam polarization described in
Section \ref{polar} has been applied. 
Moreover in this region, while the $\tilde \chi^0_1 \tilde \chi^0_2$
production cross section is typically smaller than the $\tilde \chi^0_1
\tilde \chi^0_1$
production rate, the $\tilde \chi^0_2 \tilde \chi^0_2$
production cross section can reach higher values.
At the point B belonging to this wino region and
defined as $M_1=100GeV$, $M_2=200GeV$, $\mu=600GeV$, $\tan
\beta=3$, $m_{\tilde l^{\pm}}=300GeV$
and $m_{\tilde q}=600GeV$
($m_{\tilde \chi^{\pm}_1}=189.1GeV$, $m_{\tilde \chi^0_1}=97.3GeV$,
$m_{\tilde \chi^0_2}=189.5GeV$),
the cross sections (including the beam
polarization described in Section \ref{polar})
for the allowed neutralinos productions at $\sqrt s=500GeV$ are 
$\sigma(\tilde \chi^0_1 \tilde \chi^0_1)=32.21fb$,
$\sigma(\tilde \chi^0_1 \tilde \chi^0_2)=25.60fb$
and $\sigma(\tilde \chi^0_2 \tilde \chi^0_2)=26.40fb$.\\
In the wino region and for $M_1<M_2$, the difference between
the $\tilde \chi^0_2$
and the $\tilde \chi^0_1$ masses increases with $\vert \mu \vert$
\cite{chemsin} and can reach high values. Thus, the decay
$\tilde \chi^0_2 \to \tilde \chi^0_1 + Z^0$
is often dominant. At the point B and with $m_{\tilde \nu}=450GeV$,
this channel has a branching fraction of
$B(\tilde \chi^0_2 \to \tilde \chi^0_1 Z^0)=79.9\%$.
The decay of the $Z$-boson into neutrinos has a branching fraction 
of $B(Z^0 \to \nu_p \bar \nu_p)=20.0\%$.  
In the wino region and for $M_1>M_2$, the mass difference
$m_{\tilde \chi^0_2}-m_{\tilde \chi^{\pm}_1}$ is larger than the
$W^{\pm}$ mass as long as $M_1-M_2$ is larger than $\sim 75GeV$.
In this case, the dominant channel is
$\tilde \chi^0_2 \to \tilde \chi^{\pm}_1 W^{\mp}$.
For $M_1-M_2$ smaller than $\sim 75GeV$,
$m_{\tilde \chi^0_2}-m_{\tilde \chi^{\pm}_1}$
remains large enough to allow a dominant decay of type
$\tilde \chi^0_2 \to \tilde \chi^{\pm}_1 f \bar f$, $f$ being a fermion.
As a conclusion, in the wino region the decay
$\tilde \chi^0_2 \to \tilde \chi^0_1 \bar \nu_p \nu_p$ can have a significant 
branching ratio for $M_1<M_2$.

We also consider some intermediate domains. At the point C
defined as $M_1=100GeV$, $M_2=400GeV$,
$\mu=400GeV$, $\tan \beta=3$, $m_{\tilde l^{\pm}}=300GeV$
and $m_{\tilde q}=600GeV$
($m_{\tilde \chi^{\pm}_1}=329.9GeV$, $m_{\tilde \chi^0_1}=95.5GeV$,
$m_{\tilde \chi^0_2}=332.3GeV$) 
with $m_{\tilde \nu}=450GeV$, the cross
sections (including beam polarization)
for the allowed neutralinos productions at $\sqrt s=500GeV$
are
$\sigma(\tilde \chi^0_1 \tilde \chi^0_1)=32.18fb$,
$\sigma(\tilde \chi^0_1 \tilde \chi^0_2)=6.64fb$
and $\sigma(\tilde \chi^0_1 \tilde \chi^0_3)=0.17fb$,
and the branching ratio of the $\tilde \chi^0_2$ decay into neutrinos is
$B(\tilde \chi^0_2 \to \tilde \chi^0_1 \bar \nu_p \nu_p)=0.9\%$.
The branching ratio of the decay 
$\tilde \chi^0_2 \to \tilde \chi^0_1 \bar \nu_p \nu_p$ is
small for this point C since the $\tilde \chi^0_2$ mass is large which favors
other $\tilde \chi^0_2$ decays.

At the point D given by $M_1=150GeV$, $M_2=300GeV$,
$\mu=200GeV$, $\tan \beta=3$, $m_{\tilde l^{\pm}}=300GeV$
and $m_{\tilde q}=600GeV$
($m_{\tilde \chi^{\pm}_1}=165.1GeV$, $m_{\tilde \chi^0_1}=121.6GeV$,
$m_{\tilde \chi^0_2}=190.8GeV$),
the cross sections (including beam polarization)
for the allowed neutralinos productions at $\sqrt s=500GeV$
are
$\sigma(\tilde \chi^0_1 \tilde \chi^0_1)=7.80fb$,
$\sigma(\tilde \chi^0_1 \tilde \chi^0_2)=13.09fb$,
$\sigma(\tilde \chi^0_2 \tilde \chi^0_2)=10.08fb$,
$\sigma(\tilde \chi^0_1 \tilde \chi^0_3)=35.69fb$,
$\sigma(\tilde \chi^0_2 \tilde \chi^0_3)=41.98fb$,
$\sigma(\tilde \chi^0_3 \tilde \chi^0_3)=0.02fb$
and $\sigma(\tilde \chi^0_1 \tilde \chi^0_4)=0.42fb$.
Since this point D lies in a particular region between
the higgsino region and the domain of large $\vert \mu \vert$
(or equivalently large $m_{\tilde \chi^0_3}$), the
$\tilde \chi^0_1 \tilde \chi^0_3$ and $\tilde \chi^0_2 \tilde \chi^0_3$
productions become relatively important. At the MSSM point D
and for $m_{\tilde \nu}=450GeV$,
the branching ratio of the $\tilde \chi^0_2$ decay into neutrinos is
$B(\tilde \chi^0_2 \to \tilde \chi^0_1 \bar \nu_p \nu_p)=10.6\%$.

Finally, in the domain of low charged slepton masses the $\tilde \chi^0_i
\tilde \chi^0_j$
productions are increased due to the $t$ channel $\tilde l^{\pm}$
exchange contribution.
At the point E, defined as the point B with a lower
$\tilde l^{\pm}$ mass, namely $M_1=100GeV$, $M_2=200GeV$,
$\mu=600GeV$, $\tan \beta=3$, $m_{\tilde l^{\pm}}=150GeV$ 
and $m_{\tilde q}=600GeV$, 
the polarized neutralinos pair productions rates which read as
$\sigma(\tilde \chi^0_1 \tilde \chi^0_1)=67.06fb$,
$\sigma(\tilde \chi^0_1 \tilde \chi^0_2)=57.15fb$
and $\sigma(\tilde \chi^0_2 \tilde \chi^0_2)=69.56fb$
are increased compared to the point B.\\
For $m_{\tilde \chi^0_2}>m_{\tilde l^{\pm}}$ 
or $m_{\tilde \chi^0_2}>m_{\tilde q}$,
the dominant $\tilde \chi^0_2$ decays are 
$\tilde \chi^0_2 \to \tilde \chi^0_1 \bar l_p l_p$
or $\tilde \chi^0_2 \to \tilde \chi^0_1 \bar q_p q_p$,
respectively, and the decay $\tilde \chi^0_2 \to \tilde \chi^0_1 \bar \nu_p \nu_p$ 
is typically negligible except for $m_{\tilde \chi^0_2}>m_{\tilde \nu}$. 
At the point E for $m_{\tilde \nu}>m_{\tilde \chi^0_2}$, the $\tilde \chi^0_2$
mainly decays into $\tilde \chi^0_1 \bar l_p l_p$ 
via an on shell $\tilde l_p^{\pm}$ and
the decay into neutrinos has a branching ratio of  
$B(\tilde \chi^0_2 \to \tilde \chi^0_1 \bar \nu_p \nu_p) \sim 0\%$. 
In contrast, at the point E with $m_{\tilde \nu}=160GeV$, 
$B(\tilde \chi^0_2 \to \tilde \chi^0_1 \bar \nu_p \nu_p) =32.0\%$.
Of course for $m_{\tilde l^{\pm}},m_{\tilde q}>m_{\tilde \chi^0_2}>m_{\tilde \nu}$,
the decay $\tilde \chi^0_2 \to \tilde \chi^0_1 \bar \nu_p \nu_p$ is dominant. 

$\bullet$ {\bf Sneutrino pair production}
The other source of SUSY background is the sneutrino pair production.
We discuss here the main variations of this background 
in the SUSY parameter space. First, we note that 
the $\tilde \nu_1$ pair production has the highest cross section 
among the $\tilde \nu_p \tilde \nu^*_p$ 
($p=1,2,3$ being the family index) productions
since it receives a contribution from the $t$ channel exchange of charginos.
Besides, the main contribution to the $4l+\Eslash$ signature
from the sneutrino pair production 
is the reaction $e^+ e^- \to \tilde \nu_p + \tilde \nu_p^* 
\to \tilde \chi^0_1 \nu_p + \tilde \chi^0_1 \bar \nu_p$.
Indeed, this contribution has the simplest cascade decays and furthermore
the decay $\tilde \nu_p \to \tilde \chi^0_1 \nu_p$ is favored by the phase space 
factor. Hence, we restrict the discussion of 
the $\tilde \nu_p \tilde \nu_p^*$ background to this reaction,
although all contributions from the sneutrino pair productions 
to the $4l+\Eslash$ signature are included in our analysis.

The $\tilde \nu_p$ pair production rate is reduced in the higgsino region like the 
$\tilde \chi^0_1 \tilde \chi^0_1$ production. For instance, at the point A with 
$m_{\tilde \nu}=175GeV$ the sneutrino pair production
cross section including the beam polarization effect
described in Section \ref{polar} is 
$\sigma(e^+ e^- \to \tilde \chi^0_1 \nu_p \tilde \chi^0_1 \bar \nu_p)=75.26fb$
($1.96fb$) for $p=1$ ($2,3$),
while it is 
$\sigma(e^+ e^- \to \tilde \chi^0_1 \nu_p \tilde \chi^0_1 \bar \nu_p)=501.55fb$
($10.29fb$) for $p=1$ ($2,3$)
at the point B with the same sneutrino mass.
These values of the cross sections are obtained with SUSYGEN \cite{SUSYGEN3} 
and include the spin correlations effect.
Besides, the $\tilde \nu_p$ pair production rates strongly decrease 
as the sneutrino mass increases. 
Considering once more the point B, we find that the rate is reduced to
$\sigma(e^+ e^- \to \tilde \chi^0_1 \nu_p \tilde \chi^0_1 \bar \nu_p)=232.92fb$
($4.59fb$) for $p=1$ ($2,3$) if we take now $m_{\tilde \nu}=200GeV$. \\
The branching ratio $B(\tilde \nu \to \tilde \chi^0_1 \nu)$ 
decreases as the sneutrino mass increases, since the phase space factors
associated to the decays of the sneutrino into other SUSY particles than the
$\tilde \chi^0_1$, like $\tilde \nu \to \tilde \chi^{\pm}_1 l^{\mp}$, 
increase with $m_{\tilde \nu}$. For example, at the point B 
the branching ratio $B(\tilde \nu_e \to \tilde \chi^0_1 \nu_e)$ is equal to $93.4\%$
for $m_{\tilde \nu}=175GeV$ and to $82.6\%$ for $m_{\tilde \nu}=200GeV$.

\subsection{Cuts}
\label{cuts}

\subsubsection{General selection criteria}
\label{gsc}

First, we select the events without jets containing 4 charged leptons
and missing energy.
In order to take into account the observability
of leptons at a $500GeV \ e^+ e^-$ machine, we apply the following cuts
on the transverse momentum and rapidity of all the charged leptons:
$P_t(l)>3 GeV$ and $\vert \eta(l) \vert <3$. This should
simulate the detector acceptance effects in a first approximation.
In order to reduce the \susyq background, we also demand that the number of
muons is at least equal to one. Since we consider the $\tilde \chi_1^{\pm}
\mu^{\mp}$ production, this does not affect the signal.

\subsubsection{Kinematics of the muon produced with the chargino}
\label{kin}

\begin{figure}[t]
\begin{center}
\leavevmode
\centerline{\psfig{figure=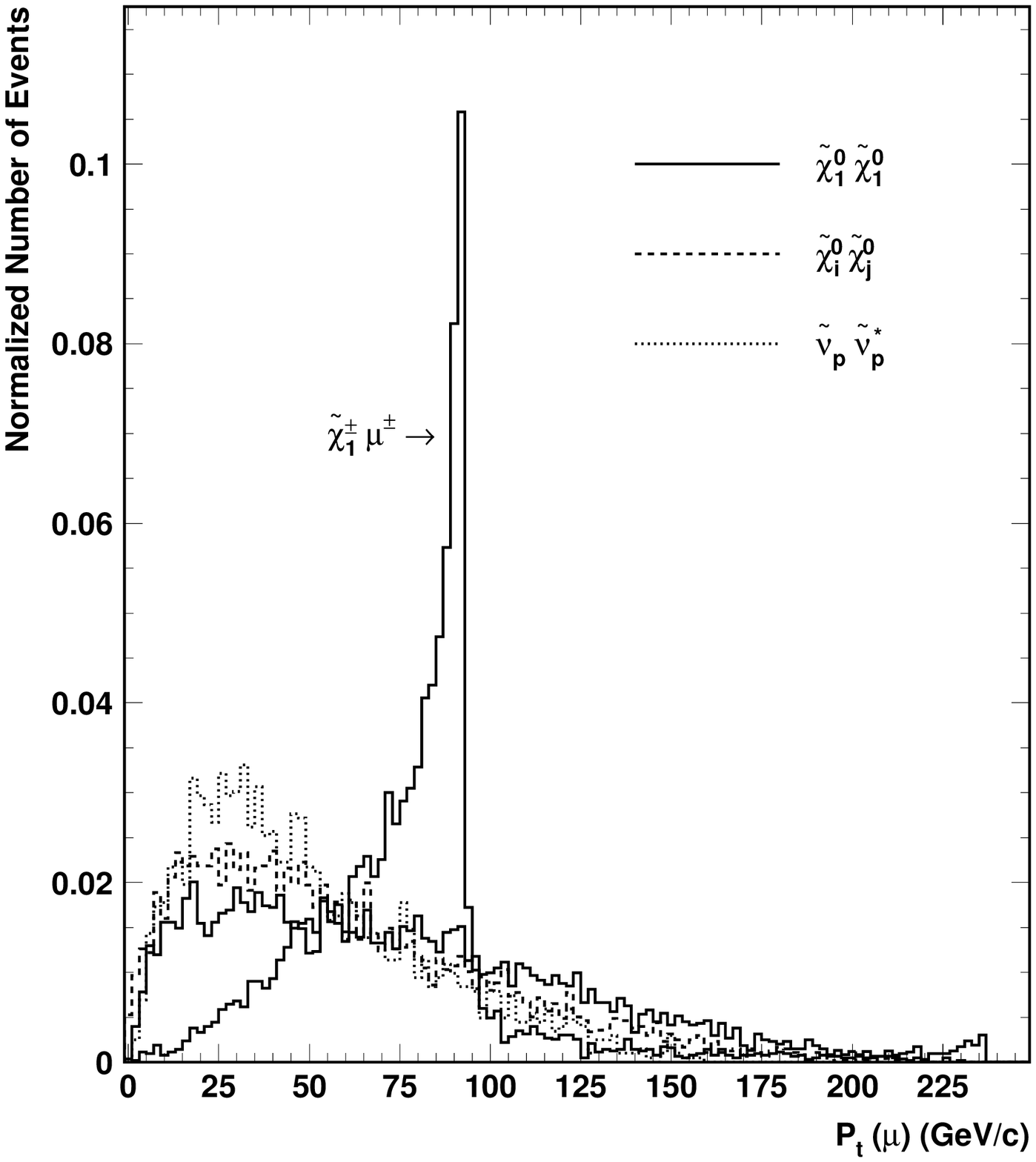,height=5.in}}
\end{center}
\caption{\footnotesize  \it
Distributions of the highest muon transverse momentum $P_t(\mu)$ (in $GeV/c$)
for the $4l + \Eslash$ events generated by
the single $\tilde \chi^{\pm}_1$ production and the SUSY background
which is
divided into the $\tilde \chi^0_1 \tilde \chi^0_1$, $\tilde \chi^0_i
\tilde \chi^0_j$ and $\tilde \nu_p \tilde \nu_p^*$ productions.
The number of events for each distribution is normalized to the unity,
the center of mass energy is fixed at $500 GeV$ and
the point A of the SUSY parameter space is considered
with $\l_{121}=0.05$ and $m_{\tilde \nu}=240GeV$.
\rm \normalsize }
\label{isr}
\end{figure}

We now discuss in details the most important cut concerning
the momentum of the muon which is produced together with the chargino.
For a negligible ISR effect, 
this muon momentum is completely determined
by the values of the chargino mass and the center of mass energy
through Eq.(\ref{enl}), as explained in Section \ref{signal}. 
Hence, some cuts on the muon momentum should
be efficient to enhance the signal-to-noise ratio
since the muon momentum distribution is perfectly peaked at a given value.
\\ For a significant ISR effect, the energy of the photon generated via the ISR 
must be taken into account. Hence, the muon energy $E(\mu)$ must be calculated
through the three-body kinematics of the reaction
$e^+ e^- \to \tilde \chi_1^{\pm} \mu^{\mp} \gamma$. Since this kinematics
depends on the angle between the muon and the photon, 
the muon energy $E(\mu)$ is not completely fixed by the SUSY parameters anymore.
Therefore, the distribution obtained experimentally of the muon momentum $P(\mu)
= ( E(\mu)^2-m_{\mu^{\pm}}^2 c^4)^{1/2}/c$ would
appear as a peaked curve instead of a Dirac peak. Although the
momentum
of the produced muon remains a good selection criteria, we have
found that in this case the transverse momentum distribution of the produced
muon was
more peaked, and thus more appropriate to apply some cuts.
We explain the reasons why in details below.\\
We have thus chosen to apply cuts on the distribution of the
muon transverse momentum instead of the whole muon momentum, even if
for a negligible ISR effect the transverse momentum distribution is not peaked as
well as the whole momentum distribution. 
The amplitude of the ISR effect depends on the SUSY mass spectrum
as we will discuss below.
\\ The cuts on the muon transverse momentum have been applied on the 
muon with the highest transverse momentum which can be identified
as the muon produced together with the chargino in the case of the signal. 
Indeed, the muon produced together with the chargino 
is typically more energetic than the 3 other leptons generated in the chargino 
cascade decay. We will come back on this point later.

In Fig.\ref{isr}, we show the distribution
of the highest muon transverse momentum $P_t(\mu)$ in the $4l + \Eslash$ events
generated by the single $\tilde \chi^{\pm}_1$ production
and the SUSY background at $\sqrt s=500 GeV$ for the point A of
the SUSY parameter space with $\l_{121}=0.05$ and $m_{\tilde \nu}=240GeV$.
For these SUSY parameters, the cross sections, polarized as in Section 
\ref{polar}, and the branching ratios are
$\sigma(\tilde \chi^{\pm}_1 \mu^{\mp})=534.39fb$, 
$B(\tilde \chi_1^{\pm} \to \tilde \chi^0_1  l_p \nu_p)=34.3\%$,
$\sigma(e^+ e^- \to \tilde \nu_p + \tilde \nu_p^* 
\to \tilde \chi^0_1 \nu_p + \tilde \chi^0_1 \bar \nu_p)=3.81fb$ and
$B(\tilde \chi^0_2 \to \tilde \chi^0_1 \bar \nu_p \nu_p)=16\%$
(see Section \ref{SUSYX} for
the values of $\sigma(\tilde \chi^0_i \tilde \chi^0_j$)).
In Fig.\ref{isr}, the cuts described in Section \ref{gsc}
have not been applied and in order to perform a separate analysis 
for each of the main considered backgrounds,
the SUSY background has been decomposed into
three components: the $\tilde \chi^0_1 \tilde \chi^0_1$, $\tilde \chi^0_i
\tilde \chi^0_j$ ($i,j=1,...,4$ with $i$ and $j$ not
equal to $1$ simultaneously) and $\tilde \nu_p \tilde \nu_p^*$ productions.
The number of $4l + \Eslash$ events 
for each of those 3 SUSY backgrounds and the single chargino
production has been normalized to the unity.
Our motivation for such a normalization is that the relative amplitudes of 
the 3 SUSY backgrounds and the single chargino
production, which depend strongly on the SUSY parameters, 
were discussed in detail in 
Sections \ref{signalX} and \ref{SUSYX}, and in this section we focus on
the shapes of the various distributions.\\ 
We see in the highest muon transverse momentum distribution 
of Fig.\ref{isr} that, as expected,
a characteristic peak arises for the single chargino production. 
Therefore, some cuts
on the highest muon transverse momentum $P_t(\mu)$
would greatly increase the signal with respect to the backgrounds.
For instance, the selection criteria suggested by the Fig.\ref{isr}
are something like 
$60GeV \stackrel{<}{\sim} P_t(\mu) \stackrel{<}{\sim} 100GeV$.
We can also observe in Fig.\ref{isr} 
that the distributions of the $\tilde \chi^0_i
\tilde \chi^0_j$ and $\tilde \nu_p \tilde \nu_p^*$ productions are
concentrated at lower transverse momentum values than 
the $\tilde \chi^0_1 \tilde \chi^0_1$ distribution.
This is due to the energy carried away by the neutrinos in the cascade
decays of the reactions (\ref{chi2}), (\ref{chi3}),
(\ref{chrp2}) and (\ref{sneu}). 
The main variation of the highest muon transverse momentum 
distributions of the SUSY backgrounds with the SUSY parameters
is the following: The SUSY backgrounds distributions
spread to larger values of the muon transverse momentum
as the $\tilde \chi^0_1$ mass increases,
since then the charged leptons coming from the
decay $\tilde \chi^0_1 \to l \bar l \nu$ reach higher energies.
\\ Fig.\ref{isr} shows also clearly that
the peak in the transverse momentum distribution of the muon produced with
the chargino exhibits an upper limit which we note $P^{lim}_t(\mu)$.
This bound $P^{lim}_t(\mu)$ is a kinematic limit
and thus its value depends on the SUSY masses. The consequence on our
analysis is that the cuts on the muon transverse momentum are
modified as the SUSY mass spectrum is changing. Note that the kinematic
limit of the whole muon momentum depends also on the
SUSY masses, so that our choice of working with
the transverse momentum remains judicious.
For a better understanding of the analysis and in view of the study
on the SUSY mass spectrum of Section \ref{marec},
we now determine the value 
of the muon transverse momentum kinematic limit $P^{lim}_t(\mu)$
as a function of the SUSY parameters.
For this purpose we divide the discussion into 2 scenarios.

\begin{figure}[t]
\begin{center}
\leavevmode
\centerline{\psfig{figure=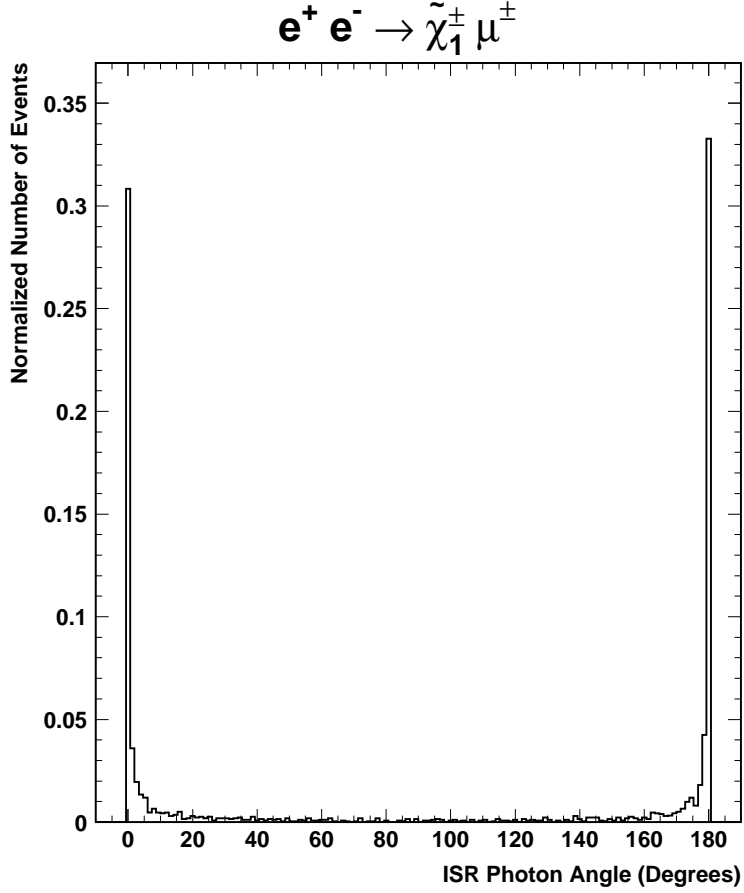,height=5.in}}
\end{center}
\caption{\footnotesize  \it
Distribution of the angle (in Degrees) between the beam axis and
the photon radiated from the initial state
of the single $\tilde \chi_1^{\pm}$ production process 
at a center of mass energy of $500GeV$ and for
the point A of the SUSY parameter
space with $m_{\tilde \nu}=240GeV$.
The number of events has been normalized to the unity.
\rm \normalsize }
\label{distribII}
\end{figure}

First, we consider the situation where $m_{\tilde \nu_{\mu}}>m_{\tilde
\chi_1^{\pm}}+m_{\mu^{\mp}}$ and $m_{\tilde \nu_{\mu}}<\sqrt s$.
In this case, the dominant
$s$ channel contribution to the three-body 
reaction $e^+ e^- \to \tilde \chi_1^{\pm}
\mu^{\mp} \gamma$, where the photon is generated by the ISR, 
can be decomposed into two levels. First,
a sneutrino is produced together with a photon in the two-body reaction
$e^+ e^- \to \gamma \tilde \nu_{\mu}$
and then the sneutrino decays as $\tilde \nu_{\mu} \to \tilde \chi_1^{\pm}
\mu^{\mp}$.
Thus, the muon energy in the center of mass of the sneutrino $E^\star(\mu)$
(throughout this article a $\star$ indicates that the variable is defined
in the center of mass of the produced sneutrino) is equal to,
\begin{eqnarray}
E^\star(\mu)=
{m^2_{\tilde \nu_{\mu}} + m^2_{\mu} - m^2_{\tilde \chi^{\pm}}
\over 2 m_{\tilde \nu_{\mu}}}.
\label{enmu}
\end{eqnarray}
The transverse momentum limit of the produced muon in the center
of mass of the sneutrino $P^{lim \star}_t(\mu)$ is given by 
$P^{lim \star}_t(\mu) =( E^{\star}(\mu)^2-m_{\mu^{\pm}}^2 c^4)^{1/2}/c$ since
$P_t^\star(\mu) \leq P^\star(\mu)$ and 
$P^\star(\mu) =( E^{\star}(\mu)^2-m_{\mu^{\pm}}^2 c^4)^{1/2}/c$.
The important point is that the sneutrino rest frame
is mainly boosted in the direction of the beam axis.
This is due to the fact that the produced photon
is mainly radiated at small angles with respect to the initial colliding
particles direction (see Fig.\ref{distribII}).
The consequence is that the transverse momentum of the 
produced muon is mainly the same in
the sneutrino rest frame and in the laboratory frame. In conclusion, the
transverse momentum limit of the produced muon in the laboratory frame is
given by $P^{lim}_t(\mu) \approx P^{lim \star}_t(\mu)
=( E^{\star}(\mu)^2-m_{\mu^{\pm}}^2 c^4)^{1/2}/c$,
$E^\star(\mu)$ being calculated through Eq.(\ref{enmu}).
We see explicitly through Eq.(\ref{enmu}) that the value of $P^{lim}_t(\mu)
\approx ( E^{\star}(\mu)^2-m_{\mu^{\pm}}^2 c^4)^{1/2}/c$ 
increases with the sneutrino mass.
\\ In this first situation, where $m_{\tilde \nu_{\mu}}>m_{\tilde
\chi_1^{\pm}}+m_{\mu^{\mp}}$
and $m_{\tilde \nu_{\mu}}<\sqrt s$, the ISR effect is large so that the
single chargino production cross section is enhanced.
The distributions of Fig.\ref{isr} are obtained for the point A
(for which $m_{\tilde \chi_1^{\pm}}=115.7GeV$) 
with $m_{\tilde \nu_{\mu}}=240GeV$
which corresponds to this situation of large ISR.
The distribution of Fig.\ref{isr} gives a value for
the transverse momentum kinematic limit of 
$P^{lim}_t(\mu)=92.3GeV/c$ which is well approximately equal to  
$( E^{\star}(\mu)^2-m_{\mu^{\pm}}^2 c^4)^{1/2}/c=92.1GeV/c$
where $E^\star(\mu)$ is calculated using Eq.(\ref{enmu}). 
\\ Based on this explanation,
we can now understand why for large ISR the transverse momentum
of the produced muon $P_t(\mu)$ is more peaked than its whole momentum
$P(\mu)$:
As we have discussed above, the muon transverse momentum $P_t(\mu)$
is controlled by the two-body kinematics of
the decay $\tilde \nu_{\mu} \to \tilde \chi_1^{\pm} \mu^{\mp}$.
For a set of SUSY parameters, 
the muon transverse momentum is thus fixed if only the
absolute value of the cosinus of the angle that makes the muon
with the initial beam direction in the sneutrino rest frame 
is given. In contrast, the whole muon momentum $P(\mu)$ 
is given by the three-body kinematics of the reaction
$e^+ e^- \to \tilde \chi_1^{\pm} \mu^{\mp} \gamma$ and thus depends on the
cosinus of the angle between the muon and the photon 
(recall that the photon
is mainly emitted along the beam axis) in the laboratory frame. 
Therefore, the muon transverse momentum $P_t(\mu)$ is
less dependent on the muon angle than the whole muon momentum $P(\mu)$,
leading thus to a more peaked distribution.

\begin{figure}[t]
\begin{center}
\leavevmode
\centerline{\psfig{figure=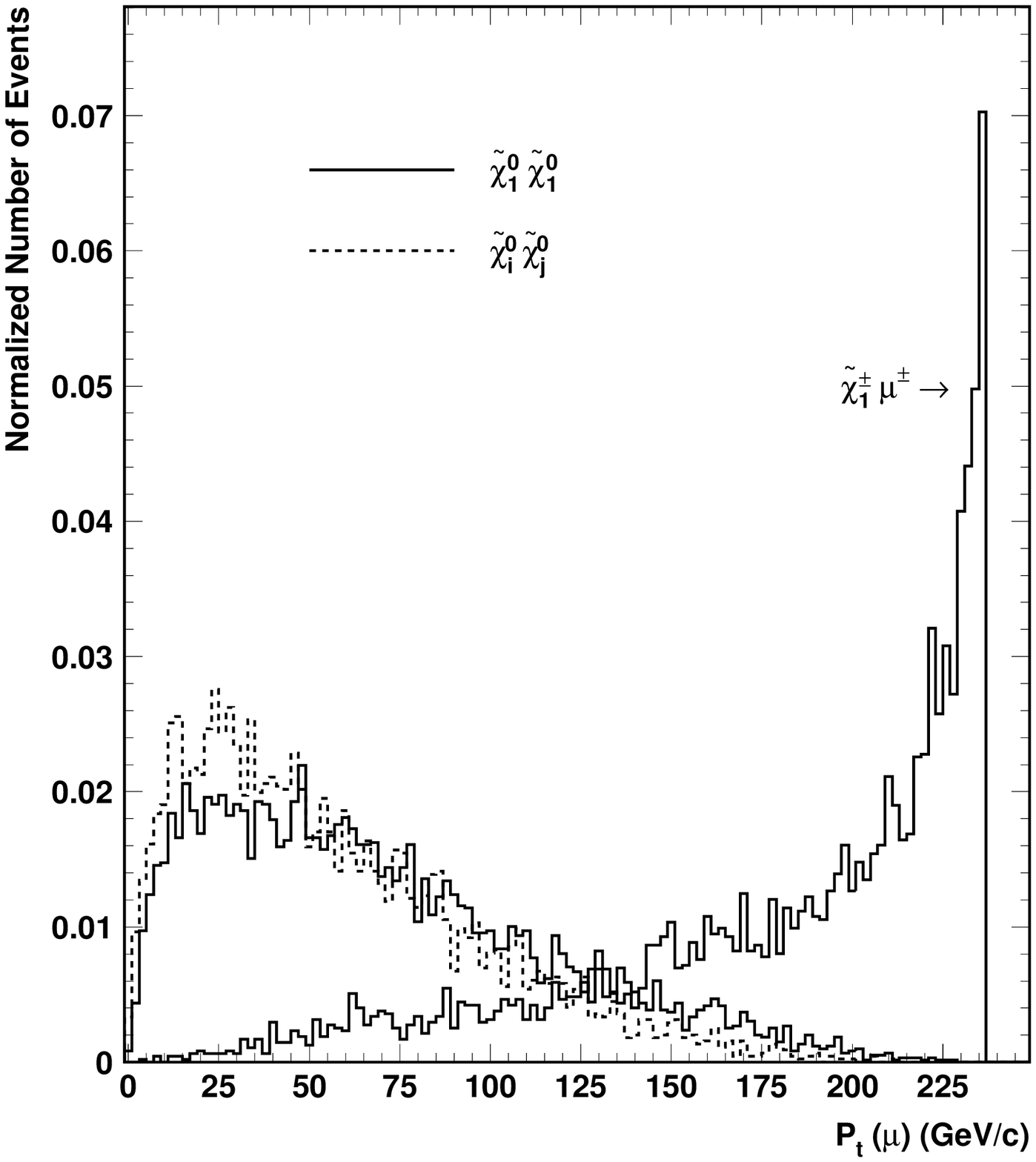,height=5.in}}
\end{center}
\caption{\footnotesize  \it
Distributions of the highest muon transverse momentum $P_t(\mu)$ (in $GeV/c$)
for the $4l + \Eslash$ events generated by
the single $\tilde \chi^{\pm}_1$ production and the SUSY background
which is
divided into the $\tilde \chi^0_1 \tilde \chi^0_1$ and $\tilde \chi^0_i
\tilde \chi^0_j$ productions.
The number of events for each distribution is normalized to the unity,
the center of mass energy is fixed at $500 GeV$ and
the point A of the SUSY parameter space is considered
with $\l_{121}=0.05$ and $m_{\tilde \nu}=550GeV$.
\rm \normalsize }
\label{noisr}
\end{figure}

We consider now the scenario where $m_{\tilde \nu_{\mu}}<m_{\tilde
\chi_1^{\pm}}+m_{\mu^{\mp}}$ or
$m_{\tilde \nu_{\mu}}>\sqrt s$. In such a situation,
the kinematics is different since the sneutrino
cannot be produced on shell.
The single chargino production $e^+ e^- \to \tilde \chi_1^{\pm}
\mu^{\mp} \gamma$ can thus not occur through the two-body
reaction $e^+ e^- \to \gamma \tilde \nu_{\mu}$. In fact in this situation,
the energy of the radiated photon becomes negligible so that the muon energy
$E(\mu)$ is given in a good approximation by
the two-body kinematics formula of Eq.(\ref{enl}).
The kinematic limit of the muon transverse momentum is related to
this energy through 
$P^{lim}_t(\mu) = P(\mu) =( E(\mu)^2-m_{\mu^{\pm}}^2 c^4)^{1/2}/c$.
\\ In this second scenario where $m_{\tilde \nu_{\mu}}<m_{\tilde
\chi_1^{\pm}}+m_{\mu^{\mp}}$ or $m_{\tilde \nu_{\mu}}>\sqrt s$, 
the ISR effect is negligible
and the single chargino production rate is thus not increased.
In Fig.\ref{noisr}, we show the transverse momentum distribution
(without the cuts of Section \ref{gsc}) 
of the produced muon in such a situation, 
namely at $\sqrt s=500 GeV$ for the point A of
the SUSY parameter space with $\l_{121}=0.05$ and $m_{\tilde
\nu_{\mu}}=550GeV$. For these SUSY parameters, 
the cross sections, polarized as in Section 
\ref{polar}, and the branching ratios are
$\sigma(\tilde \chi^{\pm}_1 \mu^{\mp})=91.05fb$, 
$B(\tilde \chi_1^{\pm} \to \tilde \chi^0_1  l_p \nu_p)=33.9\%$ and
$B(\tilde \chi^0_2 \to \tilde \chi^0_1 \bar \nu_p \nu_p)=15.7\%$
(see Section \ref{SUSYX} for
the values of $\sigma(\tilde \chi^0_i \tilde \chi^0_j$).
We check that the value of the muon transverse
momentum kinematic limit $P^{lim}_t(\mu)=236.7GeV/c$  
obtained from the distribution of Fig.\ref{noisr} 
is well approximately equal to 
$( E(\mu)^2-m_{\mu^{\pm}}^2 c^4)^{1/2}/c=236.6GeV/c$
where $E(\mu)$ is obtained from Eq.(\ref{enl}).

At this stage, we can make a comment on the variation 
of the muon transverse momentum
distribution associated to the 
single $\tilde \chi^{\pm}_1$ production with
the value of the transverse momentum limiy $P^{lim}_t(\mu)$. 
As can be seen by comparing Fig.\ref{isr} and Fig.\ref{noisr},
this distribution is more peaked for small values of
$P^{lim}_t(\mu)$ due
to an higher concentration of the distribution 
at low muon transverse momentum values.
This effect is compensated by the fact that at low muon transverse
momentum values the SUSY background is larger, as can be seen in
Fig.\ref{isr}.

The determination of $P^{lim}_t(\mu)$ performed in this section
allows us to verify
that the muon produced with the chargino is well often
the muon of highest transverse momentum. Indeed,
we have checked that in the two situations of large
(point A with $m_{\tilde
\nu_{\mu}}=240GeV$) and negligible (point A with $m_{\tilde
\nu_{\mu}}=550GeV$) ISR effect,
the calculated values of $P^{lim}_t(\mu)$ for the produced muon
were consistent with the values of $P^{lim}_t(\mu)$ obtained with
the highest muon transverse momentum distributions
(Fig.\ref{isr} and Fig.\ref{noisr}).
Therefore, the identification of the produced muon with the muon of
highest transverse momentum is correct for the two MSSM points considered
above.
The well peaked shapes of the highest muon transverse momentum distributions
for the point A with
$m_{\tilde \nu_{\mu}}=240GeV$ and $m_{\tilde
\nu_{\mu}}=550GeV$ are another confirmation of the validity of this
identification.
Nevertheless, one must wonder what is the domain of validity of this
identification.
In particular, the transverse momentum of the produced muon can be small
since $P^{lim}_t(\mu)$ which is determined via Eq.(\ref{enl}) or
Eq.(\ref{enmu}) can reach low values.
As a matter of fact, $P^{lim}_t(\mu)$ can reach small values
in the scenario where $m_{\tilde \nu_{\mu}}>m_{\tilde \chi_1^{\pm}}+m_{\mu^{\mp}}$
and $m_{\tilde \nu_{\mu}}<\sqrt s$
for $m_{\tilde \nu_{\mu}}$ close to $m_{\tilde \chi^{\pm}_1}$
(see Eq.(\ref{enmu})),
as well as in the scenario where 
$m_{\tilde \nu_{\mu}}<m_{\tilde \chi_1^{\pm}}+m_{\mu^{\mp}}$ or
$m_{\tilde \nu_{\mu}}>\sqrt s$
for $\sqrt s$ close to $m_{\tilde \chi^{\pm}_1}$ (see Eq.(\ref{enl})).
Moreover, the leptons generated in the cascade decay initiated by the
$\tilde \chi_1^{\pm}$
become more energetic, and can thus have larger transverse momentum,
for either larger $\tilde \chi_1^0$ masses or higher mass differences
between the $\tilde \chi_1^{\pm}$ and $\tilde \chi_1^0$.
We have found that for produced muons of small transverse momentum
corresponding to $P^{lim}_t(\mu) \approx 10 GeV$ the identification
remains valid for neutralino masses up to $m_{\tilde \chi_1^0} \approx
750GeV$ and mass differences up to
$m_{\tilde \chi_1^{\pm}}-m_{\tilde \chi_1^0} \approx 750GeV$.

\subsection{Discovery potential}
\label{reach}

\begin{figure}[t]
\begin{center}
\leavevmode
\centerline{\psfig{figure=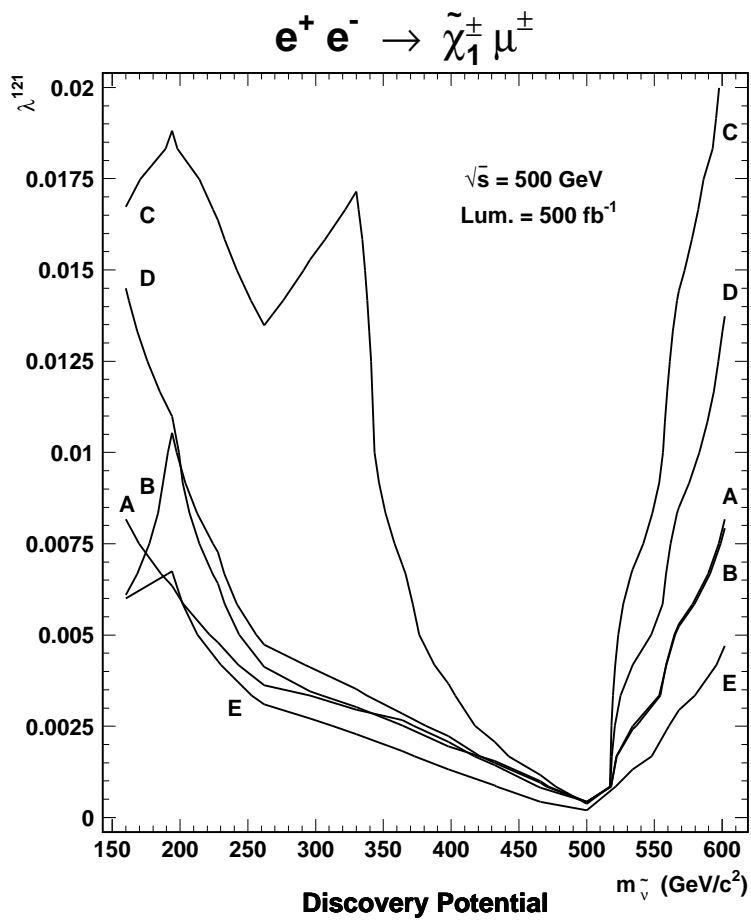,height=5.in}}
\end{center}
\caption{\footnotesize  \it
Discovery potential at the $5 \sigma$ level
in the plane $\l_{121}$ versus $m_{\tilde \nu}$ (in $GeV/c^2$) 
for the points A, B, C, D, E
of the SUSY parameter space (see text) at a center of mass energy of
$500GeV$ and assuming a luminosity of ${\cal L}=500 fb^{-1}$.
The domains of the SUSY parameter space situated above the curves
correspond to $S / \sqrt B \geq 5$ where $S$ is the $4l+\Eslash$ signal 
generated by the single $\tilde \chi^{\pm}_1$ production
and $B$ is the $R_p$-conserving SUSY background.
\rm \normalsize }
\label{reach1}
\end{figure}

In Fig.\ref{reach1}, we present exclusion plots at the $5 \sigma$ level
in the plane $\l_{121}$ versus $m_{\tilde \nu_{\mu}}$ based on the 
study of the $4l + \Eslash$ final state for several points
of the SUSY parameter space and at a center of mass energy of
$\sqrt s=500GeV$ assuming a luminosity of ${\cal L}=500 fb^{-1}$
\cite{Tesla}. The considered signal and backgrounds are respectively the
single $\tilde \chi^{\pm}_1$ production and
the pair productions of all the neutralinos and sneutrinos. 
The curves of Fig.\ref{reach1} correspond to
a number of signal events larger than 10. 
An efficiency of $30\%$ has been assumed for the reconstruction  
of the tau-lepton from its hadronic decay. 
The experimental LEP limit on the sneutrino mass $m_{\tilde \nu}>78GeV$
\cite{DELPHIb} has been respected.
We have included the ISR effects as well as the effects of the
polarization described in Section \ref{polar},
assuming an electron (positron) polarization efficiency of $85 \%$  ($60
\%$). The cuts described in Section \ref{cuts} have also been applied. 
In particular, we have
applied some cuts on the transverse momentum of the produced muon. Since
the value of the muon transverse momentum depends on the sneutrino 
and chargino masses (see Section \ref{kin}), 
different cuts have been chosen
which were appropriate to the different values
of $m_{\tilde \nu_{\mu}}$ and $m_{\tilde \chi_1^{\pm}}$ 
considered in Fig.\ref{reach1}.

We can observe in Fig.\ref{reach1} that
the sensitivity on the $\l_{121}$ \cc typically increases as
the sneutrino mass approaches the resonance point 
$m_{\tilde \nu}=\sqrt s =500GeV$. 
This is due to the ISR effect which increases 
the single chargino production rate as discussed in Section \ref{signalX}.
While the ISR effect is significant in the domain 
where $m_{\tilde \nu_{\mu}}>m_{\tilde \chi_1^{\pm}}+m_{\mu^{\mp}}$
and $m_{\tilde \nu_{\mu}}<\sqrt s$, it is small
for $m_{\tilde \nu_{\mu}}<m_{\tilde \chi_1^{\pm}}+m_{\mu^{\mp}}$ or
$m_{\tilde \nu_{\mu}}>\sqrt s$, the reason being that in 
the former case the single chargino production occurs through
the production of an on shell sneutrino as explained in Section \ref{kin}.
This results in a decrease of the sensitivity on the $\l_{121}$ coupling 
at $m_{\tilde \nu_{\mu}} \stackrel{>}{\sim} \sqrt s$ and 
$m_{\tilde \nu_{\mu}} \stackrel{>}{\sim} m_{\tilde \chi_1^{\pm}}$
as illustrate
the various curves of Fig.\ref{reach1}. The
decrease of the sensitivity corresponding to 
$m_{\tilde \nu_{\mu}} \stackrel{>}{\sim} m_{\tilde \chi_1^{\pm}}$
occurs at larger sneutrino masses for the point 
C compared to the other SUSY points, 
since for this set of SUSY parameters  
the chargino mass is larger: $m_{\tilde \chi_1^{\pm}}=329.9GeV$. 
\\ For $m_{\tilde \nu}<m_{\tilde \chi_1^{\pm}}$ the 
decay $\tilde \chi_1^{\pm} \to \tilde \chi_1^0 l_p \nu_p$ 
becomes dominant as explained in Section \ref{signalX}. 
This results in an increase of the sensitivity on $\l_{121}$ which can
be seen for the point C between $m_{\tilde \nu} \approx 330GeV$ 
and $m_{\tilde \nu} \approx 260GeV$ and
for the points B and E between $m_{\tilde \nu} \approx 190 GeV$ 
and $m_{\tilde \nu} \approx 160GeV$. 
\\ Additional comments must be made concerning 
the exclusion curve obtained for the point C:
The decrease of sensitivity in the interval 
$260GeV \stackrel{>}{\sim} m_{\tilde \nu} \stackrel{>}{\sim} 200GeV$  
comes from the increase of the
sneutrino pair production cross section (see Section \ref{SUSYX}),
while the increase of sensitivity in the range
$200GeV \stackrel{>}{\sim} m_{\tilde \nu} \stackrel{>}{\sim} 160GeV$ 
is due to an increase of the
single chargino production rate 
which receives in this domain an important $t$ channel contribution.
The significant sensitivity on the $\l_{121}$ coupling 
obtained for the point C in the interval
$330GeV \stackrel{>}{\sim} m_{\tilde \nu} \stackrel{>}{\sim} 160GeV$ 
emphasizes the importance of the off-resonance contribution
in the single chargino production study at linear colliders.
For $m_{\tilde \nu}>500GeV$, the sensitivity obtained with the point C
is weak with respect to the other SUSY points due to the high chargino
mass which suppresses the signal cross section. 
\\ We also see in Fig.\ref{reach1} that 
there are no important differences between the exclusion curves
obtained for the SUSY points belonging to the higgsino region (point A), 
the wino region (point B) and the intermediate domain (point D).
The reason is the following.
The single chargino production has a cross section which decreases
as going from the wino region to the 
higgsino region (see Section \ref{signalX}). 
However, this is also true for the $\tilde \nu_p \tilde \nu_p^*$ and most of
the $\tilde \chi_i^0 \tilde \chi_j^0$ productions (see Section \ref{SUSYX}). 
Therefore, the sensitivities on the $\l_{121}$ coupling 
obtained in the higgsino and wino region are of the same order. 
\\ Finally, we discuss the sensitivity obtained for the point E
which is defined as the point B but with a smaller charged slepton mass.
Although the neutralino pair production rate is larger for the point E
than for the point B (see Section \ref{SUSYX}),
the sensitivity obtained on $\l_{121}$ 
is higher for the point E in all the considered range of sneutrino mass. 
This is due to the branching ratio 
$B(\tilde \chi_1^{\pm} \to \tilde \chi_1^0 l_p \nu_p)$ 
which becomes important at the point E due to the hierarchy 
$m_{\tilde \chi_1^{\pm}}>m_{\tilde l^{\pm}}$
(see Section \ref{signalX}).

We mention that the sensitivity on the $\l_{121}$ coupling 
tends to decrease as $\tan \beta$ increases 
(for $sign(\mu)>0$) and is weaker
for $sign(\mu)<0$ (with $\tan \beta=3$) due to the evolution of the
single chargino production rate which was described in Section \ref{signalX}.
However, the order of magnitude of the sensitivity on $\l_{121}$ 
found in Fig.\ref{reach1} remains correct for either large $\tan \beta$,
as for instance $\tan \beta=50$, or negative $\mu$.

Let us make some concluding remarks. We see in Fig.\ref{reach1} that
the sensitivity on the $\l_{121}$ coupling reaches values typically of 
order $10^{-4}$ at the sneutrino resonance.
We also observe that for each SUSY point the $5 \sigma$ limit on the 
$\l_{121}$ coupling remains more stringent than
the low-energy limit at $2 \sigma$, 
namely $\l_{121}<0.05 \ (m_{\tilde e_R} / 100GeV)$ \cite{Bhatt},
over an interval of the sneutrino mass of $\Delta m_{\tilde \nu} 
\approx 500 GeV$ around the $\tilde \nu$ pole.
Therefore, the sensitivities on the SUSY parameters obtained via 
the single chargino production analysis at linear colliders would 
greatly improve the 
results derived from the LEP analysis (see Section \ref{intro}).
Besides, the range of $\l_{121}$ \cc values investigated at linear colliders
through the single chargino production analysis (see Fig.\ref{reach1}) 
would be complementary to the sensitivities obtained 
via the displaced vertex analysis (see Eq.(\ref{LSP})).

\subsection{SUSY mass spectrum}
\label{marec}

\subsubsection{Lightest chargino and sneutrino}
\label{lightest}

First, the sneutrino mass can be determined 
through the study of the $4l+\Eslash$ final state by 
performing a scan on the
center of mass energy in order to find the value of $\sqrt s$
at which hold the peak of the cross section
associated to the sneutrino resonance.
The accuracy on the measure of the sneutrino mass should be of order
$\delta m_{\tilde \nu} \sim \sigma_{\sqrt s}$, where $\sigma_{\sqrt s}$ is
the root mean square spread in center of mass energy
given in terms of the beam resolution $R$ by \cite{muon2},
\begin{eqnarray}
\sigma_{\sqrt s} = (7 MeV) ({R \over 0.01 \%}) ({\sqrt s \over 100GeV}).
\label{beam}
\end{eqnarray}
The values of the beam resolution
at linear colliders are expected to verify $R>1\%$.

Besides, the $\tilde \chi^{\pm}_1$
mass can be deduced from the transverse momentum
distribution of the muon produced with the chargino.
The reason is that, as we have explained in Section \ref{kin}, the
value of the muon transverse momentum limit $P^{lim}_t(\mu)$ is a
function of the $\tilde \chi^{\pm}_1$ and $\tilde \nu$ masses.
In order to discuss the experimental 
determination of the chargino mass, we
consider separately the scenarios of negligible and significant
ISR effect.

In the scenario where the ISR effect is negligible 
($m_{\tilde \nu_{\mu}}<m_{\tilde \chi_1^{\pm}}+m_{\mu^{\mp}}$ or $m_{\tilde
\nu_{\mu}}>\sqrt s$),
the value of the muon transverse momentum
limit $P^{lim}_t(\mu)$ is equal to
$( E(\mu)^2-m_{\mu^{\pm}}^2 c^4)^{1/2}/c$, $E(\mu)$ being calculated
in a good approximation with Eq.(\ref{enl}),
as described in Section \ref{kin}.
Since Eq.(\ref{enl}) gives $E(\mu)$ as a function of the center of mass
energy and the chargino mass, the experimental value of $P^{lim}_t(\mu)$
would allow to determine the $\tilde \chi_1^{\pm}$ mass. \\
We now estimate the accuracy on the chargino mass measured through this
method. In this method,
the first source of error comes from the fact that to calculate the value
of $E(\mu)$ in $P^{lim}_t(\mu)=( E(\mu)^2-m_{\mu^{\pm}}^2 c^4)^{1/2}/c$,
we use the two-body kinematics formula of
Eq.(\ref{enl}) so that the small ISR effect is neglected.
In order to include this error, we rewrite the transverse momentum kinematic
limit as
$P^{lim}_t(\mu) \pm \delta P^{lim}_t(\mu)=( E(\mu)^2-m_{\mu^{\pm}}^2
c^4)^{1/2}/c$,
$E(\mu)$ being calculated using Eq.(\ref{enl}).
By comparing the value of $E(\mu)$ calculated with Eq.(\ref{enl}) and the
value of
$P^{lim}_t(\mu)$ obtained from the transverse momentum distribution, we have
found that
$\delta P^{lim}_t(\mu) \sim 1 GeV$.
One must also take into account the experimental error on
the measure of the muon transverse momentum expected at linear colliders
which is given by:
$\delta P^{exp}_t(\mu) / P_t(\mu)^2 \leq 10^{-4} (GeV / c)^{-1} $
\cite{Tesla}.
Hence, we take $\delta P^{lim}_t(\mu) \sim 1 GeV + P^{lim}_t(\mu)^2 
10^{-4}
(GeV / c)^{-1}$. The experimental error on the muon transverse momentum
limit
depends on the value of $P^{lim}_t(\mu)$ itself and thus on the SUSY
parameters
and on the center of mass energy. However,
we have found that in the single chargino production analysis at $\sqrt
s=500GeV$,
the experimental error on the muon transverse momentum
limit never exceeds $\sim 5 GeV$.
Another source of error in the measure of the $\tilde \chi_1^{\pm}$ mass
is the root mean square spread in center of mass energy $\sigma_{\sqrt s}$
which is given by Eq.(\ref{beam}).
\\ For instance, at the point A with $m_{\tilde \nu}=550GeV$
and an energy of $\sqrt s=500GeV$,
the muon transverse momentum distribution, which is shown in Fig.\ref{noisr},
gives a value for the transverse momentum limit of $P^{lim}_t(\mu)=236.7GeV/c$. 
This value leads through
Eq.(\ref{enl}) to a chargino mass of $m_{\tilde \chi_1^{\pm}}=115.3
\pm 29.7GeV$ taking into account the different sources of error and
assuming a beam resolution of $R=1\%$ at linear colliders.
Here, the uncertainty on the chargino mass is important due to the large value 
of $P^{lim}_t(\mu)$ which increases the error $\delta P^{lim}_t(\mu)$.

In the scenario where the ISR effect is important
($m_{\tilde \nu_{\mu}}>m_{\tilde \chi_1^{\pm}}+m_{\mu^{\mp}}$ and $m_{\tilde
\nu_{\mu}}<\sqrt s$), the three-body kinematics of the reaction
$e^+ e^- \to \tilde \chi_1^{\pm} \mu^{\mp} \gamma$ leaves the muon energy
and thus the whole muon momentum $P(\mu) = ( E(\mu)^2-m_{\mu^{\pm}}^2
c^4)^{1/2}/c$ unfixed by the SUSY parameters, 
since the muon energy depends also on the angle between 
the muon and the photon. Hence, $P(\mu)$ cannot bring
any information on the SUSY mass spectrum. In contrast,
the experimental measure of the muon transverse momentum limit
$P^{lim}_t(\mu)$ would give the $\tilde \chi_1^{\pm}$
mass as a function of the $\tilde \nu_{\mu}$ mass
since $P^{lim}_t(\mu) \approx ( E^{\star}(\mu)^2-m_{\mu^{\pm}}^2
c^4)^{1/2}/c$ (see Section \ref{kin})
and $E^\star(\mu)$ is a function of $m_{\tilde \chi_1^{\pm}}$ and $m_{\tilde
\nu_{\mu}}$ (see Eq.(\ref{enmu}).
Assuming that the $\tilde \nu_{\mu}$ mass has been deduced from
the scan on the center of mass energy mentioned above,
one could thus determine the $\tilde \chi_1^{\pm}$ mass. \\
Let us evaluate the degree of precision in the measure of $m_{\tilde
\chi_1^{\pm}}$ through this calculation.
First, we rewrite the transverse momentum limit as
$P^{lim}_t(\mu) \pm  \delta P^{lim}_t(\mu)=
( E^{\star}(\mu)^2-m_{\mu^{\pm}}^2 c^4)^{1/2}/c$ to take into account
the error $\delta P^{lim}_t(\mu)$ coming from the fact that the
emission angle of the radiated photon with
the beam axis is neglected (see Section \ref{kin}). 
By comparing the value of $E^\star(\mu)$ calculated with Eq.(\ref{enmu}) 
and the value of
$P^{lim}_t(\mu)$ obtained from the transverse momentum distribution, we have
found that $\delta P^{lim}_t(\mu) \sim 1 GeV$.  
In order to consider the experimental error on $P^{lim}_t(\mu)$, we take
as before $\delta P^{lim}_t(\mu) \sim 1 GeV + P^{lim}_t(\mu)^2 
10^{-4} (GeV / c)^{-1}$.
One must also take into account the error on the sneutrino mass.
We assume that this mass has been preliminary determined up to
an accuracy of $\delta m_{\tilde \nu} \sim \sigma_{\sqrt s}$.
\\ For example, the transverse momentum distribution of Fig.\ref{isr},
which has been obtained for the point A 
with $m_{\tilde \nu}=240GeV$ and $\sqrt s=500GeV$,
gives a value for the transverse momentum limit of  
$P^{lim}_t(\mu)=92.3GeV$. The chargino mass derived from this value 
of $P^{lim}_t(\mu)$ via
Eq.(\ref{enmu}) is $m_{\tilde \chi_1^{\pm}}=115.3\pm 5.9GeV$, if 
the various uncertainties are considered and assuming that
$\sigma_{\sqrt s} \sim
3.5GeV$ which corresponds to $R=1\%$ and $\sqrt s=500GeV$ 
(see Eq.(\ref{beam})).


Hence, the $\tilde \chi_1^{\pm}$ mass can be determined
from the study of the peak in
the muon transverse momentum distribution
in the two scenarios of negligible and significant ISR effect.
The regions in the SUSY parameter space where
a peak associated to the signal can be observed over the SUSY background
at the $5 \sigma$ level are shown in Fig.\ref{reach1}.
These domains indicate for which SUSY parameters
$P^{lim}_t(\mu)$ and thus 
the chargino mass can be determined, if we assume
that as soon as a peak is observed in
the muon transverse momentum distribution
the upper kinematic limit of this peak
$P^{lim}_t(\mu)$ can be measured.

\subsubsection{Heaviest chargino}

The single $\tilde \chi^{\pm}_2$ production
$e^+ e^- \to \tilde \chi^{\pm}_2 \mu^{\mp}$ is also
of interest at linear colliders. This reaction, when kinematically open,
has a smaller phase space factor than the single $\tilde \chi^{\pm}_1$
production. Hence,
the best sensitivity on the $\l_{121}$ 
coupling is obtained from the study of the
single $\tilde \chi^{\pm}_1$ production.
However, for sufficiently large values of the $\l_{121}$ coupling,
the $e^+ e^- \to \tilde \chi^{\pm}_2 \mu^{\mp}$ reaction would
allow
to determine either the $\tilde \chi^{\pm}_2$ mass or a relation between
the $\tilde \chi^{\pm}_2$ and $\tilde \nu$ masses.
As described in Section \ref{lightest},
those informations could be derived from the upper kinematic limit
$P^{lim}_t(\mu)$ of the peak associated 
to the single $\tilde \chi^{\pm}_2$ production
observed in the muon transverse momentum distribution.
Indeed, the method presented in the study of 
the single $\tilde \chi^{\pm}_1$ production 
remains valid for the single $\tilde \chi^{\pm}_2$ 
production analysis.

\begin{figure}[t]
\begin{center}
\leavevmode
\centerline{\psfig{figure=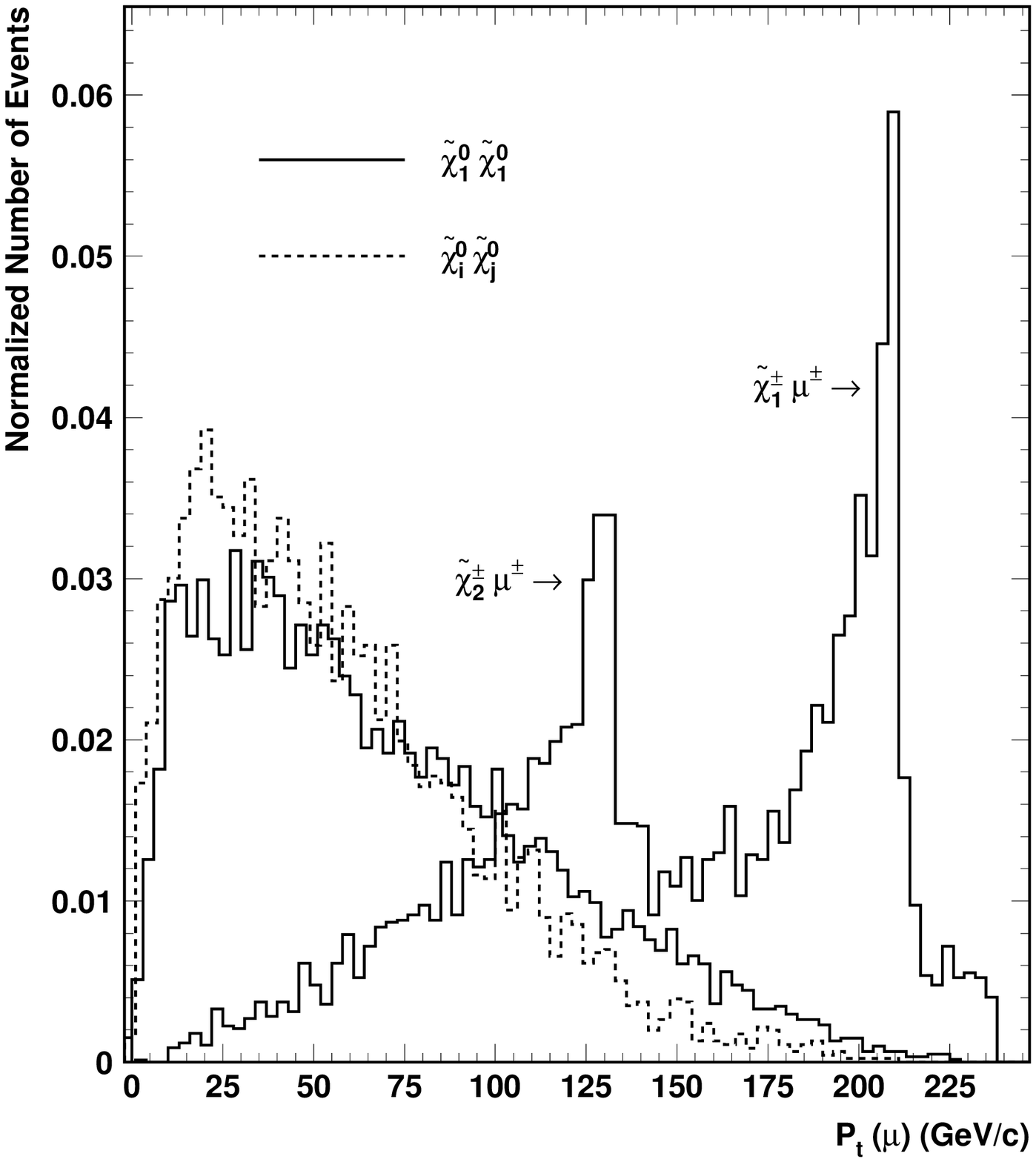,height=5.in}}
\end{center}
\caption{\footnotesize  \it
Distributions of the highest muon transverse momentum $P_t(\mu)$ (in $GeV/c$)
for the $4l + \Eslash$ events generated by
the single chargino productions
($\tilde \chi^{\pm}_1 \ + \ \tilde \chi^{\pm}_2$) 
and the SUSY background which is
divided into the $\tilde \chi^0_1 \tilde \chi^0_1$ and $\tilde \chi^0_i
\tilde \chi^0_j$ productions.
The number of events for each distribution is normalized to the unity,
the center of mass energy is fixed at $500 GeV$ and
the point A of the SUSY parameter space is considered
with $\l_{121}=0.05$ and $m_{\tilde \nu}=450GeV$.
\rm \normalsize }
\label{isr2}
\end{figure}

The simultaneous determination of the $\tilde \chi^{\pm}_1$ and 
$\tilde \chi^{\pm}_2$ masses is possible since
the peaks in the muon transverse momentum distribution
corresponding to the single $\tilde \chi^{\pm}_1$ and 
$\tilde \chi^{\pm}_2$ productions can be distinguished and identified.
In order to discuss this point,
we present in Fig.\ref{isr2} the muon transverse momentum 
distribution for the $4l + \Eslash$ events generated by
the reactions $e^+ e^- \to \tilde \chi^{\pm}_1 \mu^{\mp}$ and
$e^+ e^- \to \tilde \chi^{\pm}_2 \mu^{\mp}$ 
and by the SUSY background at the MSSM point A 
(for which $m_{\tilde \chi_1^{\pm}}=115.7GeV$ and $m_{\tilde
\chi_2^{\pm}}=290.6GeV$) with
$\l_{121}=0.05$ and $m_{\tilde \nu}=450GeV$.
The cuts described in Section \ref{gsc} have not been applied.
In this Figure, we observe that the peak associated to
the single 
$\tilde \chi^{\pm}_2$ production appears at smaller values
of the transverse momentum
than the peak caused by the single
$\tilde \chi^{\pm}_1$ production.
This is due to the hierarchy of the chargino masses. Indeed,
the chargino masses enter
the formula of Eq.(\ref{enmu}) which gives the values of 
$P^{lim}_t(\mu) \approx
( E^{\star}(\mu)^2-m_{\mu^{\pm}}^2 c^4)^{1/2}/c$
for the point A with $m_{\tilde \nu}=450GeV$. 
The same difference between
the two peaks is observed in the scenario where the values of
$P^{lim}_t(\mu) \approx ( E(\mu)^2-m_{\mu^{\pm}}^2 c^4)^{1/2}/c$ 
are calculated using the formula
of Eq.(\ref{enl}), namely for $m_{\tilde \nu_{\mu}}<m_{\tilde
\chi_1^{\pm}}+m_{\mu^{\mp}}$ and  $m_{\tilde \nu_{\mu}}<m_{\tilde
\chi_2^{\pm}}+m_{\mu^{\mp}}$, or $m_{\tilde \nu_{\mu}}>\sqrt s$.
Fig.\ref{isr2} also illustrates the fact
that the peaks associated to the single
$\tilde \chi^{\pm}_1$ and $\tilde \chi^{\pm}_2$ productions can be
easily identified thanks to their relative heights.
The difference between the heights of the two peaks is due to the relative 
values of the cross sections and branching ratios which read 
for instance with 
the SUSY parameters of Fig.\ref{isr2} as
(including the beam polarization described in Section \ref{polar}):
$\sigma(\tilde \chi^{\pm}_1 \mu^{\mp})=620.09fb$, 
$\sigma(\tilde \chi^{\pm}_2 \mu^{\mp})=605.48fb$, 
$B(\tilde \chi_1^{\pm} \to \tilde \chi^0_1  l^{\pm} \ \Eslash)=33.9\%$ and
$B(\tilde \chi_2^{\pm} \to \tilde \chi^0_1  l^{\pm} \ \Eslash)=10.8\%$.

\subsection{Extension of the analysis to different center of mass energies}

In this section, we comment on a similar study of the
reaction $e^+ e^- \to \tilde \chi^{\pm} \mu^{\mp}$ 
based on the $4l+\Eslash$ events at center of mass energies
different from $\sqrt s=500GeV$.

First, the muon transverse momentum distributions depend mainly on the
relative values of the center of mass energy and of the various SUSY masses,
so that
the discussion on the cuts given in Section \ref{cuts} still hold for
different energies.
The reconstruction of the $\tilde \chi^{\pm}_{1,2}$ and
$\tilde \nu$ masses through the methods exposed in
Section \ref{marec} is possible at any center of mass energy.
We thus discuss in this section the sensitivity on the $\l_{121}$
coupling that would be 
obtained at other center of mass energies than $\sqrt s=500GeV$.

Similarly, the values of the branching ratios are function of the SUSY mass spectrum,
as shown in Sections \ref{signalX} and \ref{SUSYX},
and thus do not change if the center of mass energy is modified.

Besides, the amplitude of the ISR effect on the signal
cross section depends on the relative values of 
$m_{\tilde \nu}$, $m_{\tilde \chi^{\pm}_1}$ and $\sqrt s$
(see Section \ref{kin}). The shapes of the exclusion curves 
obtained at different center of mass energies would thus 
be similar to the shapes of the exclusion plots presented 
in Fig.\ref{reach1} for same relative values of 
$m_{\tilde \nu}$, $m_{\tilde \chi^{\pm}_1}$ and $\sqrt s$.

However, the cross sections of the SUSY backgrounds, namely 
the $\tilde \nu_p \tilde \nu^*_p$ and $\tilde \chi^0_i
\tilde \chi^0_j$ productions, depend strongly on the 
center of mass energy which determines the phase space factors 
of the superpartners pair productions.

Therefore, the sensitivity on the $\l_{121}$ coupling tends to 
decrease (increase) at higher (smaller) center of mass energies
due to the increase (decrease) of the SUSY background rates.

\subsection{Study based on the $3l+2jets+\Eslash$ final state}
\label{hadron}

In this Section, we would like to emphasize the interest of the
$3l+2jets+\Eslash$ final state for the study of the reaction 
$e^+ e^- \to \tilde \chi^{\pm} \mu^{\mp}$ at linear colliders.

First, the $3l+2jets+\Eslash$ final state is generated by the
decay $\tilde \chi^{\pm} \to \tilde \chi^0_1 d_p u_{p'}$ 
which has a larger branching ratio than the decay
$\tilde \chi^{\pm} \to \tilde \chi^0_1  l_p \nu_p$
for the hierarchy $m_{\tilde \nu},m_{\tilde l^{\pm}},
m_{\tilde q}>m_{\tilde \chi^{\pm}_1}$,
as mentioned in Section \ref{signalX}.
\\ Secondly, 
the \sm background of the $3l+2jets+\Eslash$ signature
is the $WWZ$ production. The rate of the
$3l+2jets+\Eslash$ production 
from the $e^+ e^- \to WWZ $ reaction is
$1.5fb$ ($0.5fb$) at $\sqrt s=500GeV$ ($350GeV$)
including the cuts $\vert \eta (l)\vert<3$, $P_t(l)>10GeV$
and neglecting the ISR \cite{Godbole}.
This background can be further reduced as explained in Section
\ref{SMback}. Besides, the $3l+2jets+\Eslash$ signature
has no $R_p$-conserving SUSY background
if one assumes that the LSP is the $\tilde \chi^0_1$
and that the single dominant \rpv coupling is $\l_{121}$.
Indeed, the pair productions of SUSY particles typically
lead to final states wich contain at least 4 charged
leptons due to the decay of the two LSP's
through $\l_{121}$ as $\tilde \chi^0_1 \to l \bar l \nu$.
\\ Therefore, the sensitivity on the $\l_{121}$ 
coupling obtained from the study of the
single chargino production based on the 
$3l+2jets+\Eslash$ final state should be greatly
enhanced with respect to the analysis 
of the $4l+\Eslash$ signature.

The $3l+2jets+\Eslash$ final state is also attractive from the mass
reconstruction point of view. Indeed, the full decay chain
$\tilde \chi^{\pm}_1 \to \tilde \chi^0_1 d_p u_{p'}$,
$\tilde \chi^0_1 \to l \bar l \nu$ can be reconstructed.
The reason is that, since the muon produced together with the chargino in
the reaction
$e^+ e^- \to \tilde \chi^{\pm}_1 \mu^{\mp}$ 
can be identified (see Section \ref{kin})
the origin of each particle in the final state can 
be known. First, the $\tilde \chi^0_1$ 
mass can be measured with the distribution of
the invariant mass of the two charged leptons coming from the $\tilde
\chi^0_1$ decay:
The value of the $\tilde \chi^0_1$ mass is readen at the upper endpoint of
this distribution. Similarly, the upper endpoint 
of the invariant mass distribution of the two jets coming
from the chargino decay gives the mass difference $m_{\tilde
\chi^{\pm}_1}-m_{\tilde \chi^0_1}$.
Since $m_{\tilde \chi^0_1}$ has already been determined from the
dilepton invariant mass,
we can deduce from this mass diffference the $\tilde \chi^{\pm}_1$ mass.
\\ Besides, the $\tilde \nu$ and $\tilde \chi^{\pm}_{1,2}$ 
masses can be measured as explained
in Section \ref{marec}.
\\ Hence, in the case of a non-vanishing $\l$ coupling with $\tilde
\chi^0_1$ as the LSP,
the combinatorial background for the $\tilde \chi^0_1$
mass reconstruction from the study of the single
chargino production based on the $3l+2jets+\Eslash$ final state is expected
to be greatly reduced
with respect to the analysis based on the pair production
of SUSY particles \cite{Atlas}, due to the better
identification of the charged leptons coming from the $\tilde \chi^0_1$
decay.
Furthermore, while the $\tilde \chi^{\pm}_{1,2}$ and $\tilde \nu$
masses reconstructions are possible
via the single chargino production, these 
reconstructions appear to be more difficult
with the pair production of SUSY particles.
Indeed, the $\tilde \chi^{\pm}$ and $\tilde \nu$
masses reconstructions from the superpartner pair production
are based on the $\tilde \chi^0_1$ reconstruction, and
moreover in the pair prodution 
the $\tilde \chi^{\pm}$ or $\tilde \nu$
decays lead to either an additional uncontroled missing energy
or an higher number of charged particles (or both)
in the final state with respect to the single chargino 
production signature.

\subsection{Study of the single chargino production
through various \rpv couplings}

The single chargino production at muon colliders
$\mu^+ \mu^- \to \tilde \chi^{\pm} l^{\mp}_m$ 
(see Fig.\ref{graphe}) would allow
to study the $\l_{2m2}$ coupling, $m$ being equal to either 
$1$ or $3$ due to the antisymmetry of the $\l_{ijk}$ couplings.

The study of the $\l_{212}$ coupling would be essentially identical
to the study of $\l_{121}$.
Only small modifications would enter the analysis:
Since an electron/positron would be produced together with the chargino,
one should require that the final state contains at least one electron
instead of one muon as in the $\l_{121}$ case (Section \ref{gsc}).
Besides, one should study the distribution of the highest electron
transverse
momentum instead of the highest muon transverse momentum. 
The particularities of the muon colliders 
would cause other differences in the study:
First, the analysis would suffer an additional background
due to the $\mu$ decays in the detector. Secondly,
large polarization would imply sacrifice in luminosity
since at muon colliders this is achieved by keeping only the
larger $p_z$ muons emerging from the target \cite{muon}. Therefore,
one expects an interesting sensitivity
on the $\l_{212}$ coupling 
from the single $\tilde \chi^{\pm}_1$ production at muon
colliders
although this sensitivity should not be as high
as in the study of the $\l_{121}$ 
coupling at linear colliders. \\
Besides, the $\mu^+ \mu^- \to \tilde \chi^{\pm} e^{\mp}$ reaction occuring
through
$\l_{212}$ would allow to determine the $\tilde \chi^0_1$, $\tilde \chi^{\pm}_1$,
$\tilde \chi^{\pm}_2$ and $\tilde \nu$ masses
as described in Sections \ref{marec} and \ref{hadron} 
for the single chargino production
at linear colliders. The main difference would be the following.
The accuracy in the determination of the sneutrino mass
performed through a scan over the center of mass energy
would be higher at muon colliders than at linear colliders.
The reason is the high beam resolution $R$ expected
at muon colliders (see Section \ref{lightest}).
For instance, at muon colliders the beam resolution should reach $R \sim 0.14\%$
for a luminosity of ${\cal L}=10fb^{-1}$ and
an energy of $\sqrt s=300-500GeV$ \cite{muon}.

The production of a single chargino together with a tau-lepton
occurs through the $\l_{131}$ coupling at linear colliders and
via $\l_{232}$ at muon coliders.
Due to the $\tau$-decay, the transverse momentum
distribution of the produced $\tau$-lepton
cannot be obtained with a good accuracy.
The strong cut on the transverse momentum described
in Section \ref{kin} is thus difficult to apply in that case, and one
must think of other discrimination variables such as
the total transverse momentum of the event, the total missing energy,
the rapidity, the polar angle, the isolation angle,
the acoplanarity, the acollinearity
or the event sphericity. Therefore,
the sensitivities on the $\l_{131}$ and $\l_{232}$ couplings obtained
from the $\tilde \chi^{\pm} \tau^{\mp}$ production should be weaker than the
sensitivities expected for $\l_{121}$ and $\l_{212}$
respectively. Furthermore, the 
lack of precision in the $\tau$ transverse momentum distribution
renders the reconstructions of the $\tilde \chi^0_1$, $\tilde \chi^{\pm}_1$
and $\tilde \chi^{\pm}_2$ masses  
from the $\tilde \chi^{\pm} \tau^{\mp}$
production (see Sections \ref{marec} and \ref{hadron}) difficult.

\subsection{Single neutralino production}

We have seen in Sections \ref{kin} and \ref{reach} that when
the sneutrino mass is smaller than the lightest chargino mass
(case of small ISR effect),
the single chargino production cross section is greatly reduced
at linear colliders.
The reason is that in this situation the radiative return to the sneutrino
resonance allowed by the ISR is not possible anymore.
The interesting point is that for 
$m_{\tilde \chi^0_1}<m_{\tilde \nu}<m_{ \tilde \chi^{\pm}_1}$,
the radiative return to the sneutrino
resonance remains possible in the single $\tilde \chi^0_1$ production
through $\l_{1m1}$: $e^+ e^- \to \tilde \chi^0_1 \nu_m$.
Therefore, in the region 
$m_{\tilde \chi^0_1}<m_{\tilde \nu}<m_{ \tilde \chi^{\pm}_1}$ the
single $\tilde \chi^0_1$ production has a larger rate than
the single $\tilde \chi^{\pm}_1$ production and is thus attractive
to test the $\l_{1m1}$ couplings. 
\\ We give now some qualitative comments
on the backgrounds of the single neutralino production.
If we assume that the $\tilde \chi^0_1$ is the LSP,
the $\tilde \chi^0_1 \nu_m$
production leads to the $2l+\Eslash$ final state due to the \rpv decay
$\tilde \chi^0_1 \to l \bar l \nu$. This signature
is free of $R_p$-conserving SUSY background since the pair production
of SUSY particles produces at least 4 charged
leptons due to the presence of the LSP at the end of each of 
the 2 cascade decays.
The \sm background is strong as it comes from the
$WW$ and $ZZ$ productions but it can be reduced by some kinematic cuts.
\\ Therefore, in the region 
$m_{\tilde \chi^0_1}<m_{\tilde \nu}<m_{ \tilde \chi^{\pm}_1 }$,
one expects an interesting sensitivity on $\l_{1m1}$
from the single neutralino production study 
which was for instance considered in \cite{feng2}
for muon colliders.

The single $\tilde \chi^0_1$ production allows also to reconstruct
the $\tilde \chi^0_1$ mass. Indeed, 
the neutralino mass is given by the
endpoint of the distribution of the 2 leptons invariant mass
in the $2l+\Eslash$ final state generated by the single 
$\tilde \chi^0_1$ production.
The combinatorial background is extremely weak since
the considered final state contains only 2 charged 
leptons.

If the sneutrino is the LSP, when it is produced at the resonance
through $\l_{1m1}$ as $e^+ e^- \to \tilde \nu_m$,
it can only decay as $\tilde \nu_m \to e^+ e^-$
assuming a single dominant \rpv coupling constant.
Hence, in this scenario, the sneutrino resonance must be studied trhough
its effect on the Bhabha scattering \cite{KalZ,Kalept,Kalp,Kalseul}.
This conclusion also holds in the case of nearly degenerate $\tilde \nu$
and $\tilde \chi^0_1$ masses since then the decay $\tilde \nu_m \to  e^+
e^-$
is dominant compared to the decay $\tilde \nu_m \to \tilde \chi^0_1 \nu_m $.

\section{Conclusion}

The study at linear colliders of the single chargino production 
via the single dominant $\l_{121}$ coupling
$e^+ e^- \to \tilde \chi^{\pm} \mu^{\mp}$ 
is promising, due to the high luminosities and energies
expected at these colliders. 
Assuming that the $\tilde \chi^0_1$ is the LSP, the
singly produced chargino has 2 main decay channels: 
$\tilde \chi^{\pm} \to \tilde \chi^0_1 l^{\pm} \nu$ and 
$\tilde \chi^{\pm} \to \tilde \chi^0_1 u d$.

The leptonic decay of the produced chargino 
$\tilde \chi^{\pm} \to \tilde \chi^0_1 l^{\pm} \nu$ 
(through, virtual or real, sleptons and $W$-boson) leads to the clean 
$4l+\Eslash$ final state which is almost free of \sm background. 
This signature suffers a large SUSY background
in some regions of the MSSM parameter space but
this SUSY background can be controled
using the initial beam polarization and some 
cuts based on the specific kinematics of the single chargino production.
Therefore, considering a luminosity of ${\cal L}=500fb^{-1}$
at $\sqrt s=500GeV$ and assuming the largest SUSY background,
values of the $\l_{121}$ coupling smaller
than the present low-energy bound could be probed over a range of the sneutrino 
mass of $\Delta m_{\tilde \nu} \approx 500GeV$ 
around the sneutrino resonance and at the $\tilde \nu$ 
pole the sensitivity on $\l_{121}$ could reach values of order $10^{-4}$.
Besides, the $4l+\Eslash$ channel could allow 
to reconstruct the $\tilde \chi_1^{\pm}$,
$\tilde \chi_2^{\pm}$ and $\tilde \nu$ masses.

The hadronic decay of the produced chargino
$\tilde \chi^{\pm} \to \tilde \chi^0_1 u d$
(through, virtual or real, squarks and $W$-boson)
gives rise to the $3l+2j+\Eslash$ final state.
This signature, free from SUSY background,
has a small \sm background and should thus also
give a good sensitivity
on the $\l_{121}$ coupling constant.
This hadronic channel should allow to reconstruct the $\tilde \chi_1^0$,
$\tilde \chi_1^{\pm}$, $\tilde \chi_2^{\pm}$ and $\tilde \nu$ masses.

The sensitivity on the $\l_{131}$ coupling constant, obtained from 
the $\tilde \chi^{\pm} \tau^{\mp}$ production study, is expected to be 
weaker than the sensitivity found on $\l_{121}$ due to the decays of 
the tau-lepton.

\section{Acknowledgments}

The author is grateful to M. Chemtob, H. U. Martyn 
and Y. Sirois for helpful discussions and having encouraged this work. 
It is also a pleasure to thank N. Ghodbane and M. Boonekamp for 
instructive conversations.


\end{document}